\documentclass[journal]{IEEEtran}
\usepackage{graphicx}
\usepackage{epsf}
\usepackage{mathtools}
\usepackage{amssymb}
\usepackage{amsmath, amsfonts}
\usepackage{latexsym}
\usepackage{multirow}
\usepackage{multicol}
\usepackage[table]{xcolor}
\usepackage{color}

\usepackage{tabularx}
\usepackage{lipsum}

\usepackage{algorithm}
\usepackage{algorithmic}

\usepackage{mathrsfs}

\usepackage{fixltx2e}

\usepackage[mathscr]{euscript}

\usepackage{commath}
\usepackage{MnSymbol}

\usepackage{capt-of}
\usepackage{caption}
\usepackage{subcaption}

\usepackage{mathtools}
\DeclarePairedDelimiterX\setc[2]{[}{]}{\,#1 \;\delimsize\vert\; #2\,}
\DeclarePairedDelimiterX\parth[2]{(}{)}{\,#1 \;\delimsize\vert\; #2\,}

\DeclarePairedDelimiter{\ceil}{\lceil}{\rceil}
\DeclarePairedDelimiter{\floor}{\lfloor}{\rfloor}

\PassOptionsToPackage{hyphens}{url}
\usepackage[unicode=true, bookmarks=true, bookmarksnumbered=true, bookmarksopen=true, bookmarksopenlevel=1, breaklinks=false, pdfborder={0 0 0}, pdfborderstyle={}, backref=false, colorlinks=false]{hyperref}

\usepackage{theorem}
{
\theorembodyfont{\rmfamily}
\newtheorem{assumption}{Assumption}
\newtheorem{remark}{Remark}
\newtheorem{lemma}{Lemma}
\newtheorem{proposition}{Proposition}
\newtheorem{definition}{Definition}
\newtheorem{theorem}{Theorem}
}

\usepackage{epstopdf}

\usepackage{blkarray}
\newcommand{\matindex}[1]{\mbox{\scriptsize#1}}

\usepackage{soul}

\usepackage{bbm}
\usepackage{dsfont}

\definecolor{orange}{RGB}{255,127,0}
\definecolor{blue}{RGB}{0,0,255}
\definecolor{red}{RGB}{255,0,0}
\definecolor{green}{RGB}{50,160,50}
\definecolor{grey}{RGB}{125,120,125}
\definecolor{yellow}{RGB}{210,210,0}

\begin{document}
{

\title{\fontsize{22}{25}\selectfont Novel Backoff Mechanism for Mitigation of Congestion in DSRC Broadcast}

\author
{
Seungmo Kim, \textit{Member}, \textit{IEEE}, and Byung-Jun Kim

\thanks{S. Kim is with Department of Electrical and Computer Engineering, Georgia Southern University in Statesboro, GA, USA (e-mail: seungmokim@georgiasouthern.edu). B. J. Kim is with Department of Statistics at Virginia Tech in Blacksburg, VA, USA (e-mail: bjkim702@vt.edu).}
}

\maketitle

\begin{abstract}
Due to the recent re-allocation of the 5.9 GHz band by the US Federal Communications Commission (FCC), it will be inevitable that dedicated short-range communications (DSRC) and cellular vehicle-to-everything (C-V2X), two representative technologies for V2X, have to address coexistence. Especially, the distributed nature of DSRC makes it not trivial to control congestions, which makes it inefficient to disseminate a controlling signal. To this end, we propose a method to ``lighten'' a distributed V2X networking. Technically, it is to allocate the backoff counter according to the level of an accident risk to which a vehicle is temporarily exposed. We provide an analysis framework based on a spatiotemporal analysis. The numerical results show the improvement in the successful delivery of basic safety messages (BSMs).
\end{abstract}

\begin{IEEEkeywords}
V2X, 5.9 GHz, DSRC, CSMA, Backoff
\end{IEEEkeywords}

\IEEEpeerreviewmaketitle

\section{Introduction}\label{sec_introduction}
\subsubsection{Background on the 5.9 GHz Band}
In 1999, the US Federal Communications Commission (FCC) allocated the 5.9 GHz band for intelligent transportation system (ITS) applications based on dedicated short-range communications (DSRC) and adopted basic technical rules for the DSRC operations \cite{fcc_dsrc}. The DSRC is now at the stake of sharing the 5.9 GHz band with other radio technologies (RATs).

The first RAT is Wi-Fi. As suggested by the Congress in September 2015, the FCC, in its latest public notice \cite{fcc1668a1}, now seeks to refresh the record of its pending 5.9 GHz rulemaking to provide potential sharing solutions between proposed Unlicensed National Information Infrastructure (U-NII) devices and DSRC operations in the 5.9 GHz band. The current focus of the FCC's solicitation in \cite{fcc1668a1} is two-fold: (i) prototype of interference-avoiding devices for testing; (ii) test plans to evaluate electromagnetic compatibility of unlicensed devices and DSRC.

More recently, the cellular vehicle-to-everything communications (C-V2X) is seeking to operate in the 5.9 GHz band as well \cite{5gaa}. At present, only DSRC is permitted to operate in the ITS band in US, while the 5G Automotive Association (5GAA) has requested a waiver to the FCC to allow C-V2X operations in the band \cite{5gaa}. The key problem here is that C-V2X and DSRC are not compatible with each other. It means that if some vehicles use DSRC and others use C-V2X, these vehicles will be unable to communicate with each other--a scenario where the true potential of V2X communications cannot be attained.

\subsubsection{Significance of DSRC}
Nevertheless, DSRC still holds significance as the key technology enabling safety-critical applications \cite{5gno}. Europe recently mandated DSRC as the sole technology operating in the 5.9 GHz band \cite{europe}. Also, in the US, 50 state transport departments request reserving the 5.9 GHz band for transport safety \cite{dot}. However, state departments of transportation (DOTs) are experiencing confusion and inefficiency in enhancement/expansion of connected vehicle technologies, due to the FCC's indecisiveness on ``shared use'' of the 5.9 GHz band with other wireless systems—i.e., Wi-Fi and C-V2X \cite{dot}. Also, at the US DOT, Phase I of a three-phase testing plan has been completed, which investigated the DSRC’s interoperability with the other wireless systems \cite{fcc_phase1}. However, Phase II has not yet even started, which is supposed to involve basic field tests to assess the efficacy of findings from Phase I \cite{aashto}. Until these real-world tests are completed, one will not know conclusively to what extent, or whether at all, DSRC and the other technologies can operate together without interference, which has significant influence on the vehicle safety and mobility.

\subsubsection{Need for Lightening Load in DSRC}
A key requirement of V2V communication is the reliable delivery of safety messages. However, it has been found that a DSRC network can be congested even in simple traffic scenarios due to the limited bandwidth \cite{congestion}. In addition, the FCC reduced the bandwidth for DSRC from 75 MHz to 10 MHz under possible coexistence with C-V2X \cite{nprm}. As has been found from previous studies \cite{psunicnc}-\cite{verboom20}, it is not a trivial problem to achieve coexistence among dissimilar RATs not being able to coordinate with each other.

Relevant proposals modeled the crash risk and used it in enhancement of multiple access in V2X networks \cite{milcom19}\cite{access20}. Another method is the enhanced distributed channel access (EDCA). However, the complicated backoff model \cite{icc05} is not practical in the delay-constrained nature of a V2X safety-critical network. According to a latest study \cite{arxiv19}, the classical DSRC backoff model based on the distributed coordinated function (DCF) already yielded quite low probabilities of transmission within a beaconing period.

Motivated from the limitations of the currently available methods, this paper proposes a multiple access scheme that can be applied to a decentralized V2X network and can be performed at each vehicle without external support from infrastructure--e.g., roadside units (RSUs). The contributions of this paper are summarized as
\begin{itemize}
\item First, it defines a formula of a crash risk as a function of the variance of a vehicle's speed from the neighboring vehicles.
\item Second, this paper proposes a novel backoff allocation mechanism for DSRC, based on the crash risk. The mechanism features direct compatibility to the currently operated \textit{distributed} V2X networking standards--e.g., DSRC and C-V2X mode 3--due to its simple applicability, which shall be elaborated in Section \ref{sec_proposed_protocol}. The distributed nature is worth being kept because of the potential in a wide variety of applications--e.g., blockchain \cite{access19}.
\item It describes a carrier-sense multiple access (CSMA) using a more precise Markov chain taking into account a \textit{packet expiration}, which is idiosyncratically critical to DSRC unlike other IEEE 802.11-based technologies. In fact, our results show that a DSRC network is more constrained by an expiration rather than a collision over the air. The CSMA has a key benefit: once programmed, the protocol can be executed at each vehicle without any scheduling from the infrastructure.
\item In addition to the widely accepted classical metrics for providing easy understanding for audience, we also evaluate the proposed protocol based on inter-reception time (IRT) as a metric that adds further perspectives in analysis.
\end{itemize}

\section{Related Work}\label{sec_related_work}

\subsection{Markov Chain Model for IEEE 802.11p MAC}
Carrier sense multiple access with collision avoidance (CSMA/CA) for the general IEEE 802.11 has been modeled as a two-dimensional Markov chain \cite{bianchi}. However, in DSRC, the contention window size does not get doubled even upon a packet collision, which simplifies the Markov chain structure.

The key difference in our model is that for BSM broadcast in a DSRC system, the probability of decrementing a backoff counter is not 1. Also, a Markov chain has been proposed to model on the broadcast of safety messages in DSRC \cite{tcomp13}; yet it does not take into account a packet expiration (EXP) during a backoff process. Further, there was another aspect about which the model did not describe completely accurately. According to the IEEE 802.11p-2010 standard \cite{80211p10}, the fundamental access method of the IEEE 802.11 medium access control (MAC) is a distributed coordination function (DCF), which shall be implemented in all stations (STAs). Investigating the CSMA/CA written in IEEE 802.11 MAC \cite{bianchi}\cite{80211}, the probability of transmission at the state of backoff being 0 still requires the probability of $1 - \mathsf{P}_{b}$ for a transmission (where $\mathsf{P}_{b}$ denotes the probability of a slot being busy).

Also, the previous models suggested for analyzing the 802.11p beaconing have paid little attention to the varying number of contending nodes \cite{eurasip19_21} and the restricted channel access of the control channel (CCH), i.e., Channel 178 (5.885-5.895 GHz) \cite{eurasip19_16}. Since the 802.11p MAC protocol is a contention-based scheme, the joint effect of the varying number of contending nodes and the restricted channel access may lead the network to perform quite differently \cite{eurasip19}.

Given the significance of an EXP in determining the performance of a DSRC system, the models cannot be considered to completely accurately characterize the behavior of a safety message broadcast. In this work, we develop a new mathematic model that integrates those two factors.

\subsection{Performance Metric}
Motivated from limitations of classical metrics such as packet delivery rate (PDR) and latency, various other metrics have been proposed in the literature. The \textit{inter-reception time (IRT)} is a superior metric in the sense that it is able to display both successful and failed transmissions at once. A latest work proposed an algorithm that adapts the frequency of BSM broadcast according to the IRT \cite{dgist19}. When IRT becomes to exceed a threshold, the frequency of transmission is decremented. The decrement goes until it reaches the minimum. While this work tackles a significant problem of managing the IRT, which much previous work overlooks while relying only on classical metrics such as PDR and latency, it does not provide enough mathematical and analytical detail on how exactly formulate the IRT. The same limitation is observed from other simulation-based work \cite{vanet06}\cite{eucnc19}\cite{istc18}\cite{vnc17}\cite{eurecom17} and experiment-based work \cite{sage16}\cite{elsevier15}.

Besides the aforementioned metrics, some other metrics have also been used to evaluate the performance of congestion control techniques. Examples include the probability of successful reception of beacon message \cite{researchgate_27}, update delay \cite{researchgate_30} as the elapsed time between two consecutive BSMs successfully received from the same transmitter, a 95\% Euclidean cut-off error (in meters) \cite{researchgate_26}, and information dissemination rate (IDR) \cite{researchgate_33}.

\subsection{DSRC Performance Enhancement Scheme}
\subsubsection{EDCA}
There is body of work enhancing the performance of DSRC based on EDCA \cite{mdpi17}. However, we remind that this paper particularly aims to enhance the performance of DSRC broadcast by prioritizing vehicles at high crash risks. We consider that the EDCA is not enough to achieve that purpose. First, increasing the load of real time data traffic will increase the collision in the network. Thus, delay time and packet loss percentage will increase more than the requirements of quality of service (QoS) \cite{springer17}. Second, EDCA is known to be susceptible to additional load \cite{eurasip10}, which makes inadequate for a dynamic, distributed V2X network. Third, EDCA has been found not be able to support fairness when STAs in a network require diverse QoS requirements \cite{kim13}. Fourth, as shall be presented in Section \ref{sec_results}, a DSRC network is ``contention-constrained'' rather than collision-constrained due to the backoff process for CSMA. It yields an inefficiency of a too large CW, which in turn causes a very high probability of EXP. On the other hand, a too small CW can cause a collision, but it is rare in DSRC due to a large number of slots within a beaconing period--i.e., 1500 slots with 100 msec of inter-broadcast interval (IBI) and 66.7 $\mu$sec of a slot time \cite{elsevier14}. Considering the significance of safety-critical applications (indeed far more significant than delivering ``voice'' messages at a higher data rate), a more aggressive approach is needed to guarantee prioritization of vehicles with higher crash risk levels;

\subsubsection{Protocol Independent of CW}
Elaborating the last point, our desire is to design a protocol that optimizes a DSRC network regardless of CW. In the current 802.11 MAC mechanism, the CW size is dictated to take discrete values from a bounded finite set \cite{patel15}. The limitations of this constraint have been investigated \cite{tvt17}, to discover that the limitation on the CW size in the main bottleneck in a dense network. Also, the BEB scheme has been found not to provide adequate level of fairness due to little correlation between a backoff time and a CW \cite{beb13}. Furthermore, the IEEE 802.11 BEB backoff time calculation adjusts extremely rapidly \cite{beb13_19}; the node backs off quickly when a collision is detected and also reduces its backoff time to CWmin immediately upon a successful transmission. This produces a large variation in the backoff time; every new packet after a successful delivery starts with CWmin, which may be too small for a heavy network load. It can easily lead to a higher level of network congestion. For these reasons, application of the current BEB that heavily relies on adaptation of CW has been found inefficient for distributed networks \cite{access18}.

\subsubsection{Inter-RAT Coexistence in the 5.9 GHz Band}
The coexistence problem among dissimilar RATs in the 5.9 GHz band has been discussed in the literature: (i) between DSRC and Wi-Fi \cite{dissertation} and (ii) between DSRC and Wi-Fi/C-V2X \cite{arxiv19}.

Modification of CSMA was proposed for relieving bandwidth contention among vehicles within an IEEE 802.11p network \cite{secon19}. The inter-vehicle distance was selected as the factor representing the risk of a crash. A message prioritization scheme among different classes of vehicles was proposed for military vehicles over commercial ones \cite{milcom19}. The key limitation was a relatively simple model for the stochastic geometry: a single junction of two 6-lane road segments. Such an urban model may lose generality when applied to other scenarios.

In another latest work, a reinforcement learning-based approach was proposed to address the dynamicity of a V2X networking environment \cite{vtc20}. Each vehicle needs to recognize the frequent changes of the surroundings and apply them to its networking behavior, which was formulated as a multi-armed bandit (MAB) problem. The MAB-based reinforcement learning enabled a vehicle, without any assistance from external infrastructure, to (i) learn the environment, (ii) quantify the accident risk, and (iii) adapt its backoff counter according to the risk.

\section{System Model}\label{sec_system_model}
For formulation of the DSRC broadcast performance, this paper establishes the following key assumptions.

\begin{assumption}\label{assumption_ppp}
(Node Distribution as PPP). \textit{A generalized ``square'' space is assumed instead of an example road segment, for the most generic form of analysis as done in a related literature \cite{access19}. The environment represented by the system space $\mathbb{R}_{\text{sys}}^2$, which is defined on a rectangular coordinate with the width and length of $D$ m. Therein, a DSRC network is defined as a homogeneous Poisson point process (PPP), denoted by $\Phi_{\mathsf{D}}$, with the density of $\lambda\left(>0\right)$. The position of vehicle $i$ is denoted by $\mathtt{x}_{i}=\left(x_{i},y_{i}\right) \in \mathbb{R}_{\text{sys}}^2$. Note also that the PPP discussed in this paper is a \textit{stationary point process} where the density $\lambda$ remains constant according to different points in $\mathbb{R}_{\text{sys}}^2$.}
\end{assumption}

It is important to note that based on the modeling with PPP, the uniformity property of a homogeneous point process can be held \cite{daley}. That is, if a homogeneous point process is defined on a real linear space, then it has the characteristic that the positions of these occurrences on the real line are uniformly distributed. Therefore, we can assume that the DSRC vehicles are uniformly randomly scattered on the road with different values of intensities and CW values, which will be provided in Section \ref{sec_results}.

\begin{assumption}\label{assumption_four_types}
(Four Types of Packet Transmission Result). \textit{There are four possible results of a packet transmission including successful delivery (SUC) and two types of collision: synchronized transmission (SYNC) and hidden-node collision (HN) \cite{globecom18}. A SUC is a case where a packet does not undergo contention nor collision. A SYNC refers to a situation where more than one Tx's start transmission at the same time. A HN is the other type of collision, which occurs due to a hidden node.}
\end{assumption}

\begin{assumption}\label{assumption_cch}
(Consideration of BSM Broadcast in CCH). \textit{The analysis framework and result that will be presented throughout this paper are based on assumption of using the CCH only--i.e., channel 178 in the 5.9 GHz band. It means that the result has a room for improvement if the network's channel selection is expanded among the other shared channels (SCHs). As such, the results that will be demonstrated in Section \ref{sec_results} can be regarded as the worst-case, most conservative ones.}
\end{assumption}

\begin{assumption}\label{assumption_speed}
(Speed as the Representative Factor of Crash Risk). \textit{Speed has been found as the most direct indicator of a crash risk \cite{speed}. This paper assumes that each vehicle moves at speed of $v$ meters per second (m/s), which is a varied factor as shall be shown in Section \ref{sec_results}. The direction of a vehicle follows a uniform distribution in the range of $\left[0, 2\pi\right]$. A node is bounced off when reaching at the end of $\mathbb{R}^2$ in order to stay in the space, which hence keeps the total intensity the same.}
\end{assumption}

\begin{remark}\label{remark_speed_normal}
(Distribution of vehicle speed). \textit{Speeds are selected by the driver. Different drivers select different speeds, dependent upon many variables (vehicle limitations, roadway conditions, driver ability, etc.). No single speed value can accurately represent all the speeds at a certain location. A speed distribution provides that information. Operating speeds have been found to be normally distributed \cite{nhtsa09}. This is fortunate since using that premise (probability is normally distributed) allows for some straightforward calculations.}
\end{remark}

\begin{definition}\label{definition_Psi}
(Variance of speed: crash risk measurement metric). \textit{We use the variance of speed from the speed limit of the road as the metric measuring the risk of a crash. The rationale behind this is the ``easiness'' in getting a speed limit. For instance, a speed limit is an easy number to obtain in many commercial GPS applications (e.g., Google Maps). Reliance on such an easily available quantity increases the applicability of the proposed algorithm. It can be replaced with other relevant parameters: e.g., the mean speed over the neighboring vehicles.}
\begin{align}
\Psi = \sqrt{ \left(v - \text{v}_{L}\right)^2 }
\end{align}
where $\text{v}_{L}$ gives the speed limit of the road.
\end{definition}

\begin{definition}\label{definition_marked}
(Marked point process). \textit{The PPP $\Phi$ is defined as a ``marked'' point process where the speed variance, $\Psi$, is associated with each point $\mathbf{x}_i$. This mark is an independent normal random variable as seen from a point $\mathbf{x}_i$. Specifically, let point process $\Phi = \{ \mathbf{x}_i; i \in \mathbb{N} \}$ denote the locations of the nodes.}
\end{definition}
Importantly, it is assumed that the mark of a point does not depend on the location of its corresponding point in the underlying (state) space.
}

\section{Proposed Protocol}\label{sec_proposed_protocol}
As have already mentioned in Section \ref{sec_system_model}, we choose the variation of speed, denoted by $\Psi$, as the key indicator of a crash risk to a vehicle is exposed while driving. As such, the proposed protocol allocates the backoff counter according to the quantity of $\Psi$.

We wanted a more aggressive scheme to benefit ``dangerous'' vehicles than allocating a higher access category (AC) in EDCA. In essence, the performance of EDCA still depends on CW, which may be hazardous for a very urgent safety-related packet delivery.

We remind that the key metric in the proposed protocol is $\Psi$ that was presented in Proposition \ref{definition_Psi}. The main idea of the proposed protocol is to divide the distribution of $\Psi$ into multiple \textit{discrete sections} and apply different \textit{backoff allocation} patterns.

\begin{figure}[t]
\centering
\includegraphics[width = 0.9\linewidth]{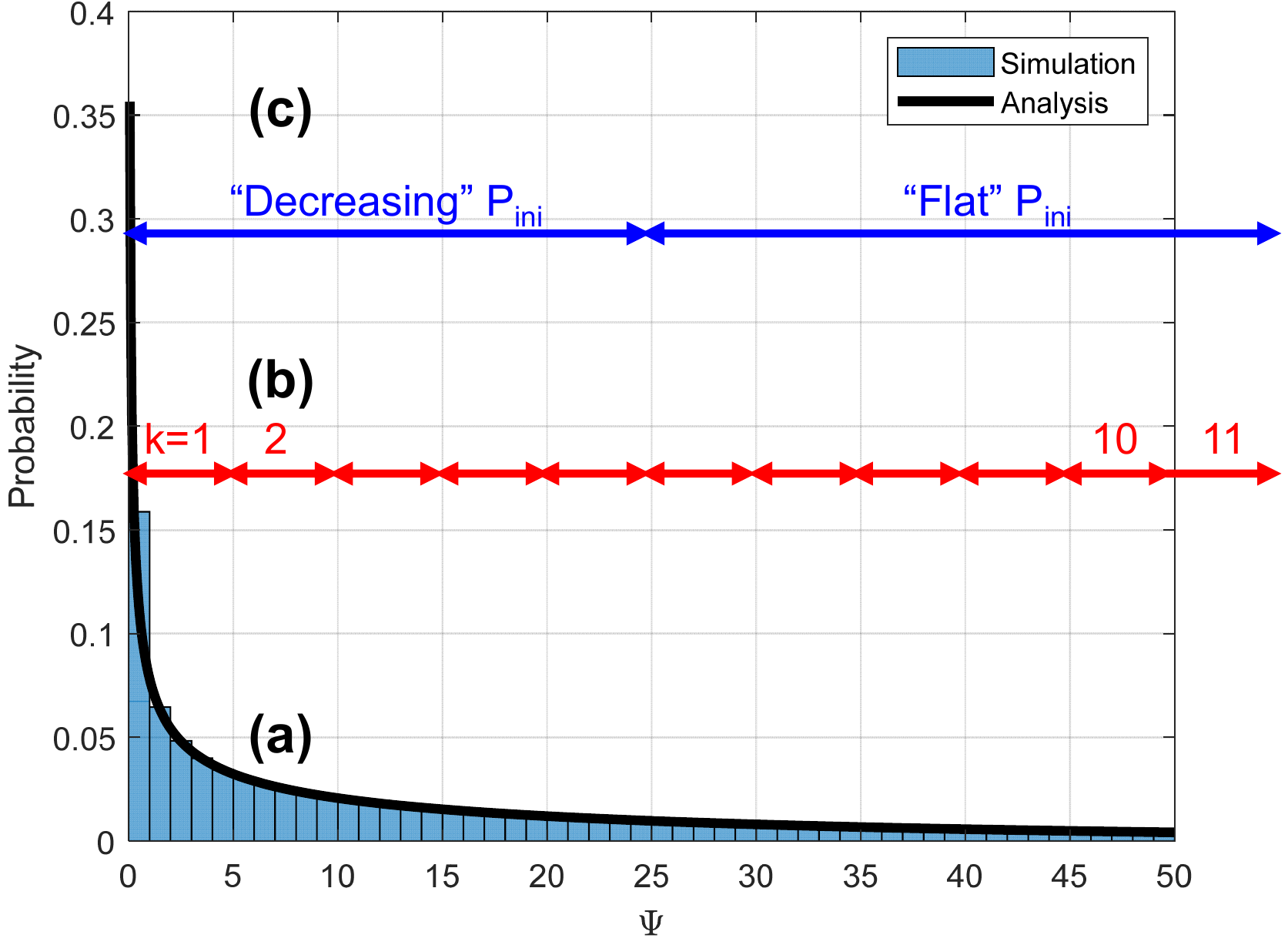}
\caption{PDF of $\Psi$ (with $X \sim \mathcal{N}\left(60, 25 \right)$): (a) Model validation between simulation and analysis of $f_{S}\left(s\right)$; (b) Categorization of $\Psi$ (with $K = 11$ and $Q=5$); and (c) An example allocation of backoff values for the proposed protocol (division of ``decreasing'' and ``flat'' at $k = \floor{K/2}$)}
\label{fig_Psi_dist}
\end{figure}

\vspace{0.1 in}
As such, we start with formulating the distribution of $\Psi$. We find the probability distribution function (PDF) as in the following Lemma:
\vspace{-0.1 in}

\begin{lemma}\label{lemma_sigma_distribution}
(Distribution of $\Psi$). \textit{As shown in Fig. \ref{fig_Psi_dist}, the speed's variance, $\Psi$, is found to follow the following distribution:}
\begin{align}
f_{\Psi}\left(\psi\right) = \frac{1}{\sqrt{2\pi}\sigma} \psi^{-\frac{1}{2}} e^{-\frac{\psi}{2\sigma^2}}, \text{~~~} \psi \ge 0
\end{align}
\end{lemma}

\textit{Proof:} Let random variable $\Psi$ be defined as $\Psi = \left( X - \mu \right)^2$ where $X \sim \mathcal{N}\left(\mu, \sigma^2 \right)$.

For $\psi < 0$, since $\Psi$ cannot have a negative value,
\begin{align}
\mathbb{P}\left( \Psi < \psi \right) = 0
\end{align}

For $\psi \ge 0$,
\begin{align}
\mathbb{P}\left( \Psi < \psi \right) &= \mathbb{P}\left( \left( X - \mu \right)^2 < \psi \right)\nonumber\\
&= \mathbb{P}\left( -\sqrt{\psi} < X - \mu < \sqrt{\psi} \right)\nonumber\\
&= \mathbb{P}\left( -\sqrt{\psi} + \mu < X < \sqrt{\psi} + \mu \right)\nonumber\\
&= F_{X}\left( \mu + \sqrt{\psi}\right) - F_{X}\left( \mu - \sqrt{\psi} \right)
\end{align}

Therefore,
\begin{align}
&f_{\Psi}\left(\psi\right)\nonumber\\
&= \frac{\text{d}}{\text{d}\psi} \left[ F_{X}\left( \mu + \sqrt{\psi}\right) - F_{X}\left( \mu - \sqrt{\psi} \right) \right]\nonumber\\
&= \frac{\text{d}}{\text{d}\psi} \Bigg[ \displaystyle \int_{-\infty}^{\mu + \sqrt{\psi}} \frac{1}{\sqrt{2\pi}\sigma} e^{\frac{-(t - \mu)^2}{2\sigma^2}} \text{d}t\nonumber\\
&\hspace{0.5 in}- \displaystyle \int_{-\infty}^{\mu - \sqrt{\psi}} \frac{1}{\sqrt{2\pi}\sigma} e^{\frac{-(t - \mu)^2}{2\sigma^2}} \text{d}t \Bigg]\nonumber\\
&= \frac{1}{\sqrt{2\pi}\sigma} \frac{\text{d}}{\text{d}\psi} \left[ \displaystyle \int_{-\infty}^{\mu + \sqrt{\psi}} e^{\frac{-(t - \mu)^2}{2\sigma^2}} \text{d}t - \displaystyle \int_{-\infty}^{\mu - \sqrt{\psi}} e^{\frac{-(t - \mu)^2}{2\sigma^2}} \text{d}t \right]\nonumber\\
&= \frac{1}{\sqrt{2\pi}\sigma} \left[ e^{-\frac{\psi}{2\sigma^2}} \frac{d}{\text{d}\psi} \left(\mu + \sqrt{\psi}\right) - e^{-\frac{\psi}{2\sigma^2}} \frac{\text{d}}{\text{d}\psi} \left(\mu - \sqrt{\psi}\right) \right]\nonumber\\
&= \frac{1}{\sqrt{2\pi}\sigma} \left[ e^{-\frac{\psi}{2\sigma^2}} \frac{1}{2} \psi^{-\frac{1}{2}} + e^{-\frac{\psi}{2\sigma^2}} \frac{1}{2} \psi^{-\frac{1}{2}} \right]\nonumber\\
&= \frac{1}{\sqrt{2\pi}\sigma} \psi^{-\frac{1}{2}} e^{-\frac{\psi}{2\sigma^2}} 
\end{align}
which completes the proof.
\hfill$\blacksquare$

\vspace{0.1 in}
To represent the crash risk, we categorize the metric, $\Psi$, into a number of regions on its PDF. The reason for this categorization is, as seen from Fig. \ref{fig_Psi_dist}, the PDF of $\Psi$ is open-ended to the right (positive side); hence, a value of $\Psi$ itself is not appropriate to be used in categorization.
\vspace{-0.1 in}

\begin{proposition}\label{proposition_sigma_categorization}
(Categorization of $\Psi$). \textit{Note that two key parameters are defined to identify a categorization: (i) the number of categories, $K$, and (ii) the step size, $Q$. The range of $\Psi$ for each category, $k$, is given by}
\begin{align}\label{eq_range_sigmav}
\left(k-1\right) Q + 1 \le \Psi \le kQ,
\end{align}
\textit{which, in turn, gives $k$ as a function of $\Psi$ as}
\begin{align}\label{eq_range_k}
k = f\left(\Psi\right) = \left[\frac{\Psi}{Q}, \frac{\left(\Psi-1\right)}{Q} + 1\right], {\rm{~~}} k \in \mathbb{Z}
\end{align}
\end{proposition}

\begin{assumption}\label{assumption_k_fixed}
(Fixed $k$ during a backoff allocation process). During a backoff process for a BSM that is shown in Fig. \ref{fig_markov}, the value of $k$ is assumed to be fixed; in other words, a vehicle does not experience a change of $\Psi$ greater than $Q$. We assume 10 Hz of the BSM generation frequency--i.e., 10 BSMs per second, which leads to 100 msec per BSM. Notice that a 100 msec is a short time in relation to the reality on the road--i.e., for a vehicle to experience a change in $\Psi$. Moreover, even if so, as an IEEE 802.11-based system, DSRC is supposed to support such a situation at ``best effort.''
\end{assumption}

\begin{figure}[t]
\centering
\includegraphics[width = \linewidth]{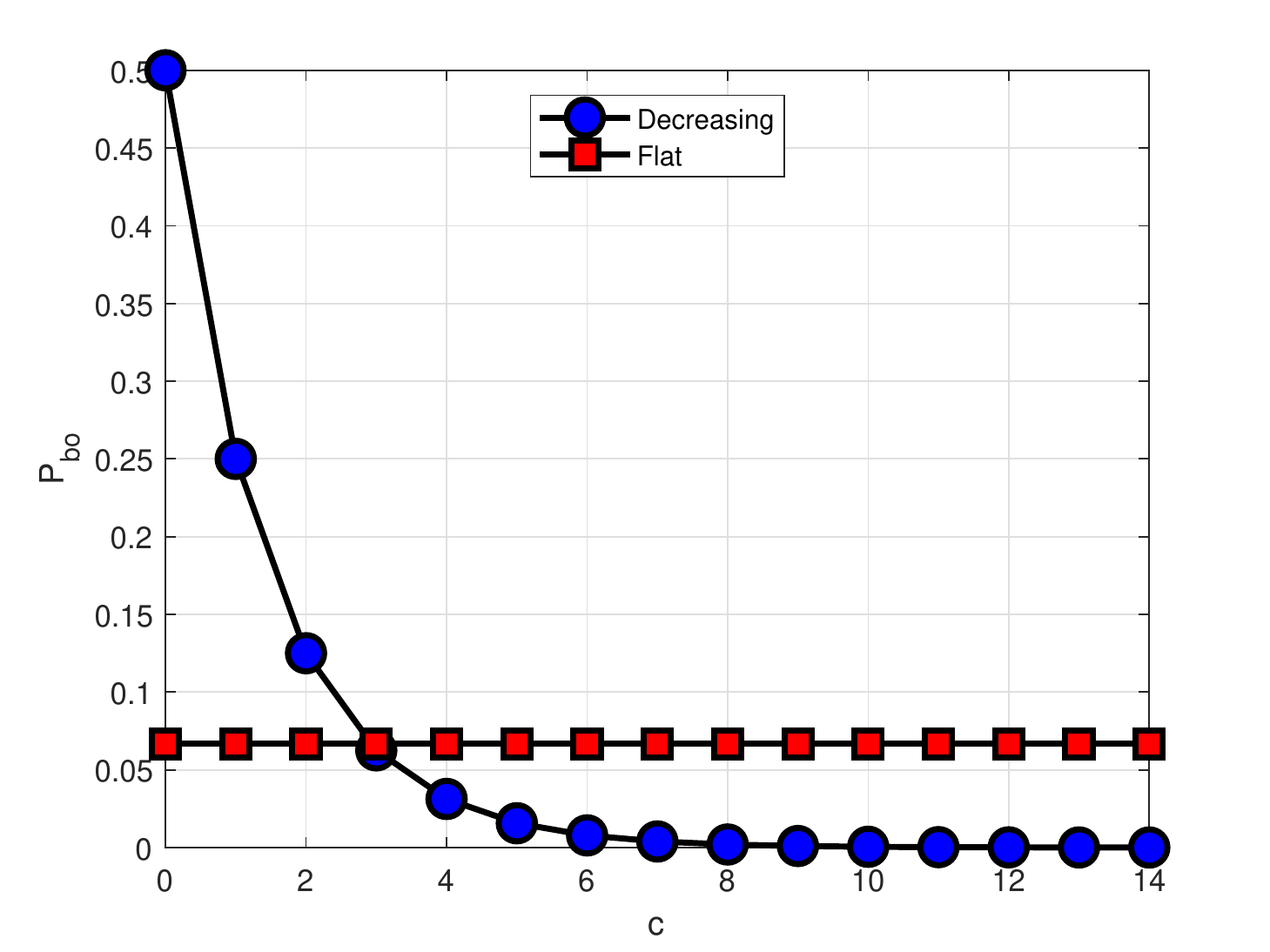}
\caption{$\mathsf{P}_{\text{ini}}$ versus $c \in \{0, 1, \cdots, \text{CW}-1\}$ (with CW $= 15$) according to the ``decreasing'' (proposed) and ``flat'' (traditional) patterns of backoff counter allocation}
\label{fig_Pini_vs_c}
\end{figure}

\vspace{0.2 in}
We define two types of backoff allocation function versus $c \in \{0, 1, \cdots, \text{CW}-1\}$:
\vspace{-0.1 in}

\begin{definition}\label{definition_Pini}
(Backoff allocation according to $\Psi$). \textit{The proposed protocol allocates $\mathsf{P}_{{\rm{bo}}}\left(c\right) \coloneqq \mathbb{P}\left[{\rm{backoff}} = c\right]$ based on the categorization of $\Psi$ into $k$, as described in Proposition \ref{proposition_sigma_categorization} and Fig. \ref{fig_Psi_dist}. The formulation is given by}
\begin{align}\label{eq_Pini}
\mathsf{P}_{\text{ini}}\left(c, k\right) &\coloneqq \mathsf{P}_{{\rm{bo}}}\left(c\right), \text{~~where~} c\in \{0, 1, \cdots, \text{CW}-1\}\nonumber\\
&= \begin{cases} r^{c+1} \text{ (``Decreasing''), } {\rm{~~}} k > \ceil{\frac{K}{2}}\\
\frac{1}{\text{CW}} \text{ (``Flat''), } {\rm{~~}} k \le \floor{\frac{K}{2}}\end{cases}
\end{align}
\textit{where term $r$ indicates the intensity of decrease in a $\mathsf{P}_{\rm{bo}}\left(c\right)$ and is defined as $r = 1/2$ to make $\sum_{c=0}^{\text{CW}-1} \mathsf{P}_{\text{ini}} = 1$, which is a mathematical requirement to make a Markov chain valid.}
\end{definition}

\vspace{-0.1 in}
More specifically, with a value of $k$ being larger than $\ceil{\frac{K}{2}}$ (i.e., the area of ``large'' $\Psi$'s, which indicates cases of greater deviations in a vehicle's speed), we allocate backoff counters in a ``decreasing'' pattern. In contrast, for $k$ being smaller than $\ceil{\frac{K}{2}}$ (i.e., the area of ``small'' $\Psi$'s, which means cases of smaller deviations in a vehicle's speed), the backoff counters are allocated in a traditional ``flat'' fashion in which the probability of any backoff counter value is equal.

Fig. \ref{fig_Psi_dist} demonstrates an example with $K=11$ and $Q=5$. On the PDF, it indicates the principle of the proposed protocol (which shall be described in Section \ref{sec_proposed_protocol}): a larger value of $k \in \{1,2,\cdots,K\}$ (which occurs with a smaller probability, $f_{S}\left(s\right)$) takes a smaller backoff value with a higher probability, $\mathsf{P}_{\text{ini}}$, as shall be given in (\ref{eq_Pini}).

\begin{remark}\label{remark_backoff_allocation}
(Interpretation of (\ref{eq_Pini})). \textit{In a ``decreasing'' backoff allocation, a node is given a higher probability for a smaller backoff value, which increases the chance of winning the medium. A ``flat'' pattern indicates backoff allocation in a uniform distribution, which is currently adopted by DSRC \cite{arxiv19}.}
\end{remark}

Analyzing Definition \ref{definition_Pini} further, we can proceed to quantify the probability that a vehicle is allocated a backoff counter following either a decreasing or a flat type function.
\vspace{-0.1 in}

\begin{proposition}\label{proposition_division_point}
(Probabilities of ``decreasing'' and ``flat'' backoff allocation functions). \textit{The proportion of nodes being allocated decreasing and flat patterns of $\mathsf{P}_{\text{ini}}$ is a predominating factor determining the performance of the proposed system. The proposed protocol determines the proportion by setting a division point in terms of $k$, to the left and the right of which are assigned for decreasing and flat patterns, respectively. This division can be formally written as}
\begin{align}\label{eq_Pdec}
\mathsf{P}_{\text{dec}} &= \displaystyle \int_{0}^{Q\ceil{\frac{K}{2}}} f_{\Psi} \left(\Psi\right)d\Psi\nonumber\\
&= \displaystyle \int_{0}^{Q\ceil{\frac{K}{2}}} \frac{1}{\sqrt{2\pi}\sigma} \psi^{-\frac{1}{2}} e^{-\frac{\psi}{2\sigma^2}} d\Psi\nonumber\\
&=\bigg[\text{erf}\left(\frac{\sqrt{\Psi}}{\sqrt{2}\sigma}\right)\bigg]_{0}^{Q\ceil{\frac{K}{2}}}\nonumber\\
&=\text{erf}\left(\frac{\sqrt{Q\ceil{\frac{K}{2}}}}{\sqrt{2}\sigma}\right)
\end{align}
\textit{and}
\begin{align}\label{eq_Pflt}
\mathsf{P}_{\text{flat}} &= \displaystyle \int_{Q\floor{\frac{K}{2}}}^{\infty} f_{\Psi} \left(\Psi\right)d\Psi\nonumber\\
&=\bigg[\text{erf}\left(\frac{\sqrt{\Psi}}{\sqrt{2}\sigma}\right)\bigg]_{Q\floor{\frac{K}{2}}}^{\infty}\nonumber\\
&=1 - \text{erf}\left(\frac{\sqrt{Q\ceil{\frac{K}{2}}}}{\sqrt{2}\sigma}\right)
\end{align}
\textit{where $\sigma$ denotes the standard deviation of random variable $\Psi$.}
\end{proposition}

\vspace{-0.1 in}
Observing Fig. \ref{fig_Psi_dist} again, the lefthand and righthand sides of the area under the PDF of $\Psi$ are $\mathsf{P}_{\text{dec}}$ and $\mathsf{P}_{\text{flat}}$, respectively.

\vspace{0.1 in}
It will be worth noticing a few remarks on the proposed protocol, which are given as follows:
\vspace{-0.1 in}
\begin{remark}\label{remark_thinning}
(Tx filtering as thinning of $\Phi$). \textit{Such a filtering of Tx vehicles by using the proposed protocol is formulated as a thinning of the PPP $\Phi$. The thinned process has a new intensity, $\tilde{\lambda} \le \lambda$. The rationale has already been mentioned in Proposition \ref{proposition_sigma_categorization}. The proposed protcol is expected to grant a higher chance of transmission for vehicles with higher crash risks. As such, when a vehicle's speed varies too much from the speed limit, the vehicle is easily able to broadcast a packet since such a large value of $\Psi$ takes a low probability as shown in Fig. \ref{fig_Psi_dist}.}
\end{remark}

\begin{remark}\label{remark_compatibilty}
(Backward compatibility of the proposed protocol). \textit{The key benefit of the proposed protocol is modification of the CSMA that is already adopted in DSRC. As such, it will achieve a higher backward compatibility and therefore an easier adoption in practice.}
\end{remark}

\begin{figure}[t]
\centering
\includegraphics[width = \linewidth]{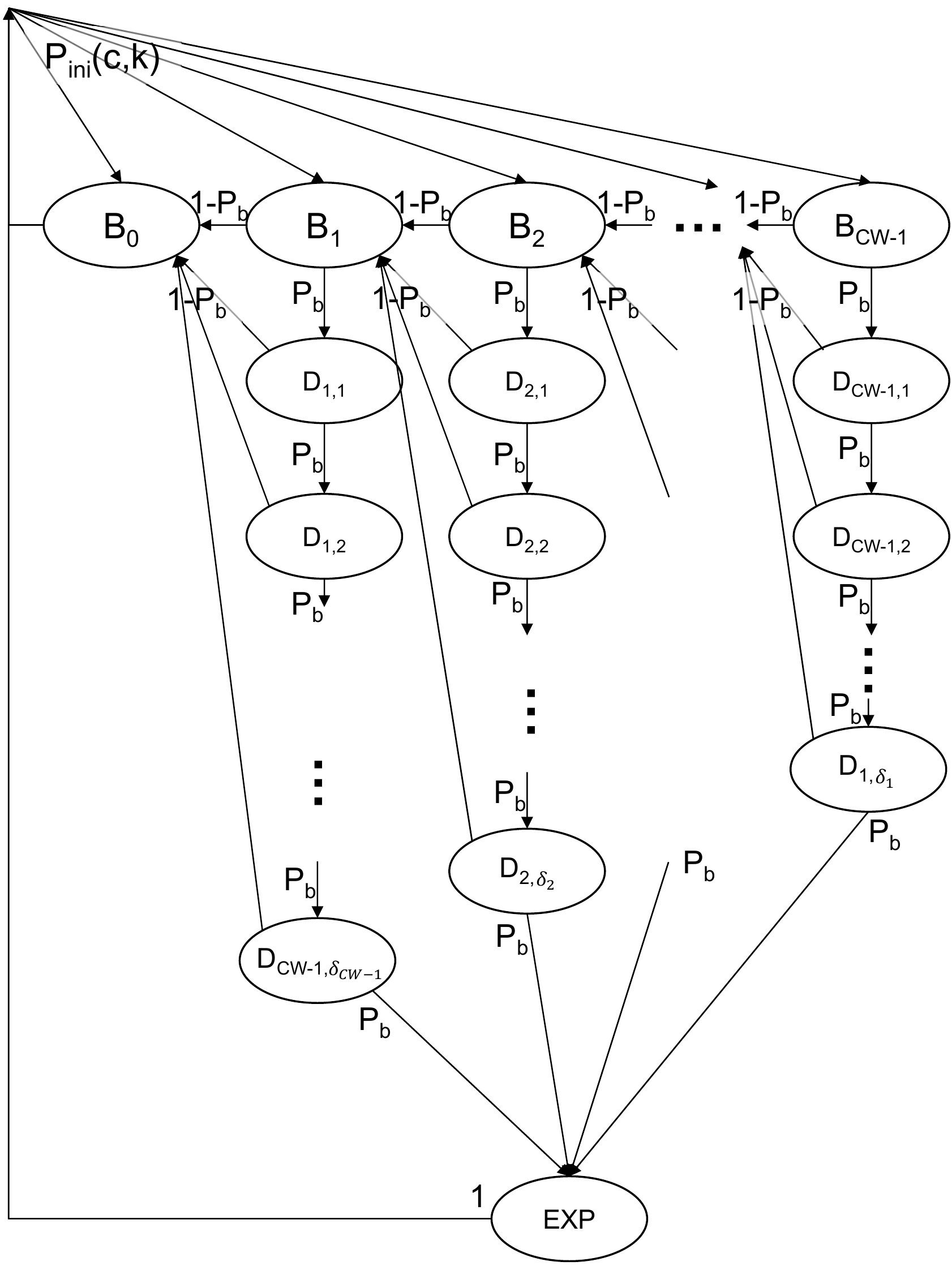}
\caption{The proposed protocol as a Markov chain}
\label{fig_markov}
\end{figure}

\section{Stochastic Analysis}\label{sec_stochastic}
For stochastic analysis, the proposed protocol operated at a vehicle is modeled as a one-dimensional Markov chain as illustrated in Fig. \ref{fig_markov}. We use fixed-point iterations to solve the model.

\subsection{Markov Chain Representation of DSRC CSMA/CA}\label{sec_markov}
To recall, in IEEE 802.11 DCF CSMA/CA \cite{80211}, if no medium activity is indicated for the duration of a particular backoff slot, then the backoff procedure shall decrement its backoff time by a slot time. If the medium is determined to be busy at any time during a backoff slot, then the backoff procedure is suspended; that is, the backoff timer shall not decrement for that slot. The medium shall be determined to be idle for the duration of a DIFS period or EIFS, as appropriate, before the backoff procedure is allowed to resume.

As another reminder, the key difference between the traditional DSRC and our proposed backoff allocation scheme lies in $\mathsf{P}_{\text{ini}}$. The traditional DSRC adopts the CSMA/CA in which the probability of being allocated a backoff counter is uniform among all the $B_{c}$'s with $c \in \left\{0, 1, \cdots, \text{CW}-1\right\}$. In comparison, the proposed scheme differentiates this probability according to the variation of the speed, $\Psi$. This makes significant difference in the probability of a packet transmission since, depending on the value of $\mathsf{P}_{b}$, it can be very difficult to make it through the Markov chain and reach $B_{0}$. Specifically, the proposed protocol allocates higher probabilities of starting the chain from $B_{c}$'s with smaller $c$'s to a vehicle with a larger $\Psi$. Especially when $\mathsf{P}_{b}$ is higher and thus challenging to propagate toward $B_{0}$, the proposed protocol will significantly prioritize the vehicles with large $\Psi$'s in the medium competition. Furthermore, as have mentioned in Section \ref{sec_introduction}, this is the first work to completely precisely display the effect of an EXP.

In the Markov chain shown in Fig. \ref{fig_markov}, $\mathsf{D}_{c,m}$ where $c \in \{0, 1, \cdots, \text{CW}-1\} \text{ and } m \in \{1, 2, \cdots, \delta_{c}\}$ denotes state of the $m$th delay at backoff state $B_{c}$. The deepest state of a delay, $\mathsf{D}_{c,\delta_{c}}$, means the last possible delay for decrement of the backoff counter from $c$ to $c-1$ before expiration of the packet.

\subsection{Probability of Transmission in a Beacon Period}\label{sec_stochastic_tau}
Let $\tau$ denote the probability that a node transmits in any slot within a beaconing period (which is composed of $L_{bcn}$ slots). In other words, $\tau$ gives the probability that a node has been able to reach $B_{0}$ in the Markov chain after going through the backoff process within a beaconing period without experiencing a packet expiration.

\begin{figure*}[t]
\begin{align}\label{eq_transition_matrix}
\mathcal{M} = 
\begin{blockarray}{ccccccccccccc}
&\matindex{$B_{0}$} & \matindex{$B_{1}$} & \cdots & \matindex{$B_{\text{CW}-1}$} & \matindex{$D_{1,1}$} & \cdots & \matindex{$D_{1,\delta_{1}}$} & \cdots & \matindex{$D_{\text{CW}-1,1}$} &\cdots & \matindex{$D_{\text{CW}-1, \delta_{\text{CW}-1}}$} & \matindex{$\text{EXP}$}\\
\begin{block}{c[cccccccccccc]}
\matindex{$B_{0}$} & \mathsf{P}_{\text{ini}}\left(0, k\right) & \mathsf{P}_{\text{ini}}\left(1, k\right) & \cdots & \mathsf{P}_{\text{ini}}\left(\text{CW}-1, k\right) & 0 & 0 & 0 & \cdots & 0 &\cdots & 0 & 0\\
\matindex{$B_{1}$} & 1 - \mathsf{P}_{b} & 0 & \cdots & 0 & \mathsf{P}_{b} & \cdots & 0 & \cdots & 0 & \cdots & 0 & 0\\
\matindex{\vdots} & \vdots & \vdots & \ddots & \vdots & \vdots & \ddots & \vdots & \ddots & \vdots & \ddots & \vdots & \vdots\\
\matindex{$B_{\text{CW}-1}$} & 0 & 0 & \cdots & 0 & 0 & \cdots & 0 & \cdots & \mathsf{P}_{b} & \cdots & 0 & 0\\
\matindex{$D_{1,1}$} & 1 - \mathsf{P}_{b} & 0 & \cdots & 0 & 0 & \cdots & 0 & \cdots & 0 & \cdots & 0 & 0\\
\matindex{\vdots} & \vdots & \vdots & \cdots & \vdots & \vdots & \ddots & \vdots & \ddots & \vdots & \cdots & \ddots & \vdots\\
\matindex{$D_{1,\delta_{1}}$} & 1 - \mathsf{P}_{b} & 0 & \cdots & 0 & 0 & \cdots & 0 & \cdots & 0 & \cdots & 0 & \mathsf{P}_{b}\\
\matindex{\vdots} & \vdots & \vdots & \cdots & \vdots & \vdots & \ddots & \vdots & \ddots & \vdots & \ddots & \vdots & \vdots\\
\matindex{$D_{\text{CW}-1,\delta_{1}}$} & 0 & 0 & \cdots & 0 & 0 & \cdots & 0 & \cdots & 0 & \cdots & 0 & \mathsf{P}_{b}\\
\matindex{$\text{EXP}$} & 0 & 0 & \cdots & 0 & 0 & \cdots & 0 & \cdots & 0 & \cdots & 0 & 1\\
\end{block}
\end{blockarray}
\end{align}
\end{figure*}

\begin{figure*}
\begin{align}\label{eq_one_step_steady_state}
&\mathsf{P}_{B_{c-1}}\nonumber\\
&{\rm{~}}= \mathsf{P}_{B_{c}} \mathbb{P}\left[ B_{c} \rightarrow B_{c-1} \right]\nonumber\\
&\stackrel{(a)}{=} \overbrace{\Big[ \underbrace{\mathsf{P}_{\text{ini}} \left(c, k\right)}_{\text{Directly from } B_{0}} + \underbrace{\mathsf{P}_{B_{c+1}} \mathbb{P}\left[ B_{c+1} \rightarrow B_{c} \right]}_{\text{Via } B_{c+1}} \Big]}^{= \mathsf{P}_{B_{c}}} \mathbb{P}\left[ B_{c} \rightarrow B_{c-1} \right]\nonumber\\
&\stackrel{(b)}{=} \bigg[ \mathsf{P}_{\text{ini}} \left(c, k\right) + \overbrace{\Big\{ \underbrace{\mathsf{P}_{\text{ini}} \left(c+1, k\right)}_{\text{Directly from } B_{0}} + \underbrace{\mathsf{P}_{B_{c+2}}\mathbb{P}\left[ B_{c+2} \rightarrow B_{c+1} \right]}_{\text{Via } B_{c+2}} \Big\}}^{= \mathsf{P}_{B_{c+1}}} \mathbb{P}\left[ B_{c+1} \rightarrow B_{c} \right] \bigg] \mathbb{P}\left[ B_{c} \rightarrow B_{c-1} \right]\nonumber\\
&{\rm{~}}= \bigg[ \mathsf{P}_{\text{ini}} \left(c, k\right) + \Big\{ \mathsf{P}_{\text{ini}} \left(c+1, k\right) + \overbrace{\Big( \underbrace{\mathsf{P}_{\text{ini}} \left(c+2, k\right)}_{\text{Directly from } B_{0}} + \underbrace{\mathsf{P}_{B_{c+2}} \mathbb{P}\left[ B_{c+3} \rightarrow B_{c+2} \right]}_{\text{Via } B_{c+2}} \Big)}^{= \mathsf{P}_{B_{c+2}}} \mathbb{P}\left[ B_{c+2} \rightarrow B_{c+1} \right] \Big\} \mathbb{P}\left[ B_{c+1} \rightarrow B_{c} \right] \bigg] \mathbb{P}\left[ B_{c} \rightarrow B_{c-1} \right]\nonumber\\
&\stackrel{(c)}{=} \bigg[ \mathsf{P}_{\text{ini}} \left(c, k\right) + \Big\{ \mathsf{P}_{\text{ini}} \left(c+1, k\right) + \overbrace{\Big( \underbrace{\mathsf{P}_{\text{ini}} \left(c+2, k\right)}_{\text{Directly from } B_{0}} + \underbrace{\mathsf{P}_{B_{c+2}} }_{\text{Via } B_{c+2}} \left( 1 - \mathsf{P}_{b} \right) \Big)}^{= \mathsf{P}_{B_{c+2}}} \left( 1 - \mathsf{P}_{b} \right) \Big\} \left( 1 - \mathsf{P}_{b} \right) \bigg] \left( 1 - \mathsf{P}_{b} \right)\nonumber\\
&\stackrel{(d)}{=} \text{The expansion is kept until reaching } \mathbb{P}\left[B_{\text{CW}-1}\right]
\end{align}
\end{figure*}

\begin{proposition}\label{proposition_transition_matrix}
(Transition matrix of the Markov chain). \textit{The transition matrix for the Markov chain (presented in Fig. \ref{fig_markov}) is denoted by $\mathcal{M}$, which is formulated in (\ref{eq_transition_matrix}) where $\delta_{c} \in \mathbb{Z}$ gives the maximum number of busy slots that a node can experience before a packet expiration (EXP) at backoff state $c$, which is formally written as}
\begin{align}\label{eq_delta_c}
\delta_{c} = L_{bcn} - l_{bcn} - c
\end{align}
\textit{with $c \in \{0, 1, \cdots, \text{CW}-1\}$. Note that $- l_{bcn}$ comes from the length of a BSM itself, while $-c$ is given from the number of backoffs that has to be decremented. It is noteworthy from (\ref{eq_delta_c}) that $\delta_{c}$ gets larger with a smaller $c$, which is translated to that a STA assigned a smaller backoff counter $c$ has a higher chance to reach $B_{0}$.}
\end{proposition}

\begin{proposition}\label{proposition_one_step_transition}
(One-step steady-state transition). \textit{Let $\mathsf{P}_{B_{c}} = \lim_{t \rightarrow \infty} \mathbb{P} \left[B(t) = B_{c}\right]$, with $B(t)$ giving a backoff counter state at time $t$ and $c \in \left\{0, 1, \cdots, \text{CW}-1\right\}$, be the stationary distribution of the chain to obtain the stead-state expressions for the Markov chain. A one-step transition between two arbitrary states (i.e., $B_{c} \rightarrow B_{c-1}$) in steady state for the Markov chain can be formulated as in (\ref{eq_one_step_steady_state}). Note from (a) that the probability of state $B_{c}$ is composed of two parts: (i) directly from $B_{0}$ and (ii) via $B_{c+1}$. As shown in (b), the same holds for $B_{c+1}$, from which one can infer the pattern of steady-state propagation for the Markov chain. Also, in (c), we substitute $\mathbb{P}\left[B_{c} \rightarrow B_{c-1}\right]$ with $1 - \mathsf{P}_{b}$ as illustrated in Fig. \ref{fig_markov}. Lastly, (d) tells that the expansion shown in (\ref{eq_one_step_steady_state}) goes until it reaches $\mathbb{P}\left[B_{\text{CW}-1}\right]$. It is important to note that state $B_{\text{CW}-1}$ has only one input: directly from $B_{0}$.}
\end{proposition}

Now, the following two propositions present the formulation of a numerical solution for $\tau$ and $\mathsf{P}_{b}$.

\begin{proposition}\label{proposition_tau}
($\tau$ as a function of $\mathsf{P}_{b}$). \textit{The probability of reaching state $B_0$ is equivalent to the probability of a packet transmission after propagating through the Markov chain \cite{bianchi}. The general one-step steady-state transition shown in Proposition \ref{proposition_one_step_transition} can be applied to obtain the probability of transition from an arbitrary state $B_{c}$ to $\mathsf{P}_{B_{0}}$. In this paper, this transition formula provides a piece for the numerical solve of $\tau$ and $\mathsf{P}_{b}$, which is given by}
\begin{align}\label{eq_tau}
\tau &:= \mathsf{P}_{B_{0}}\nonumber\\
&= \mathsf{P}_{B_{1}} \underbrace{\left(1 - \mathsf{P}_{b}\right)}_{= \mathbb{P}\left[ B_{1} \rightarrow B_{0}\right]}\nonumber\\
&= \Big(\mathsf{P}_{\text{ini}}\left(1,k\right) + \mathsf{P}_{B_{2}} \underbrace{\left(1 - \mathsf{P}_{b}\right)}_{= \mathbb{P}\left[ B_{2} \rightarrow B_{1} \right]} \Big) \left(1 - \mathsf{P}_{b}\right)\nonumber\\
&= \left(\mathsf{P}_{\text{ini}}\left(1,k\right) + \left(\mathsf{P}_{\text{ini}}\left(2,k\right) + \cdots \right)\left(1 - \mathsf{P}_{b}\right) \right) \left(1 - \mathsf{P}_{b}\right).
\end{align}
\textit{Implication of (\ref{eq_tau}) is that now $tau$ is written as a function of $\mathsf{P}_{b}$ since all the $\mathsf{P}_{\text{ini}}\left(c,k\right)$'s are known.}
\end{proposition}

\begin{proposition}\label{proposition_Pb_approx}
($\tau$ as a function of $\mathsf{P}_{b}$ and $n_{cs}$). \textit{While the Markov chain describes a node's process of going through a backoff process in CSMA/CA, $\mathsf{P}_{b}$ is determined from the dynamics among the nodes competing for the channel. More specifically, all the nodes that are located in the carrier-sense range of the tagged vehicle become the competitors for the channel. The probability of a slot being found busy can be formulated as}
\begin{align}\label{eq_Pb_approx}
\mathsf{P}_{b} &= 1 - \left( 1 - \tau \right)^{n_{cs}}
\end{align}
\textit{where $n_{cs}$ denotes the number of nodes within a given node's range of carrier sensing.}
\end{proposition}

\begin{lemma}\label{lemma_tau_Pb}
(Numerical solution for $\tau$). \textit{We compute $\tau$ and $\mathsf{P}_{b}$ via numerical solve based on simultaneous equations composed of (\ref{eq_tau}) and (\ref{eq_Pb_approx}) that have been presented in Propositions \ref{proposition_tau} and \ref{proposition_Pb_approx}, respectively. The simultaneous equations were to be solved via a sufficiently large number of iterations testing all the possible values for $\tau$ and $\mathsf{P}_{b}$.}
\end{lemma}

\subsection{Number of Slots Spent until a Transmission}\label{sec_stochastic_Nbo}
Denote by $\mathsf{N}_{\text{bo}}$ the number of time slots that has been spent for going through a backoff process. We suggest that $\mathsf{N}_{\text{bo}}$ is used as a metric that can be used to measure the performance of a network.

Notice from the Markov chain in Fig. \ref{fig_markov} that $\mathsf{N}_{\text{bo}}$ increases as $d$ increments with the probability of $\mathsf{P}_{b}$ along with the vertical direction. For instance, a STA can stay at $c$ while it experiences $k$ busy slots, which occurs with the probability of $\left(\mathsf{P}_{b}\right)^{k}$. As such, it is likely that a large number $k \rightarrow \delta_{c}$ takes a very small probability, i.e., $\left(\mathsf{P}_{b}\right)^{k} \rightarrow 0$. This yields that $\mathsf{N}_{\text{bo}}$ does not diverge too significantly. This behavior of $\mathbb{E}\left[\mathsf{N}_{\text{bo}}\right]$ can be formulated as
\begin{align}
\mathbb{E}\left[\mathsf{N}_{\text{bo}}\right] &\hspace{0.04 in}= \displaystyle \sum_{c = 1}^{\text{CW}} \sum_{d = 1}^{\delta_{c}} \mathbb{P}\left[ \left(c,d\right) \right] \mathsf{N}_{bo} \left(c,d\right)\nonumber\\
&\stackrel{(a)}{=} \displaystyle \sum_{c = 1}^{\text{CW}} \sum_{d = 1}^{\delta_{c}} \mathsf{P}_{{\rm{bo}}}\left(c\right)\left(\mathsf{P}_{b}\right)^{d} \mathsf{N}_{bo} \left(c,d\right)
\end{align}
where $\mathbb{P}\left[ \left(c,d\right) \right]$ denotes the probability that a STA is assigned the backoff counter $c$ and has additionally backed off due to $d$ busy slots. Also, $\mathsf{P}_{{\rm{bo}}}\left(c\right)$ has already been defined and discussed in Definition \ref{definition_Pini}. For instance, the typical IEEE 802.11 CSMA/CA gives $\mathsf{P}_{{\rm{bo}}}\left(c\right) = 1 /\text{CW}$ for $\forall c = 0, 1, \cdots, \text{CW}-1$.

\subsection{Packet Delivery Rate ($\mathsf{PDR}$)}\label{sec_stochastic_PDR}
Appreciating its easiness to understand, this paper adopts the probability that a packet is successfully delivered to an arbitrary receiver vehicle (or $\mathsf{PDR}$) as the key metric evaluating the proposed protocol in comparison to the original DSRC.

\begin{lemma}\label{lemma_Pexp}
(Probability of an EXP). \textit{The probability that a node is not able to transmit in any slot within a beaconing period can be formulated as}
\begin{align}
\mathsf{P}_{\text{exp}} = 1 - \tau
\end{align}
\end{lemma}

\textit{Proof:} It is significant to note that a backoff process of DSRC is ``reset'' after every beaconing period regardless of whether a packet is successfully transmitted or not. It means that the ``steady-state'' in DSRC is $t \approx L_{bcn} \times \text{(Slot time)}$ rather than $t \rightarrow \infty$ as in other IEEE 802.11 standards. As such, cases of finishing at other states including $B$'s and $D$'s occur although a separate state of EXP exists in the backoff process model as shown in Fig. \ref{fig_markov}. Recall from Fig. \ref{fig_markov} that state $\mathsf{D}_{c,\delta_{c}}$, the deepest state of a delay with backoff counter $c$, gives the last possible delay state before being drained to an EXP. It implies the possibility of staying at any other state than $B_{0}$ and EXP at the end of a beaconing period, which consequently should be added to the probability of an EXP. As a consequence, the probability of an EXP is obtained as $\mathsf{P}_{\text{exp}} = 1 - \tau$.
\hfill$\blacksquare$

Once a packet is transmitted, in order to lead to a successful delivery, no collision must occur on the packet during the transit. In other words, for a tagged vehicle, an external collision occurs unless both of the following conditions are satisfied: (1) when the tagged vehicle is transmitting, no vehicles within its carrier-sensing range delivery packets at the same time slot; and (2) when the tagged vehicle is transmitting, no hidden terminals should corrupt the tagged vehicle's packet. 

To formulate, as just have mentioned, there are two types of collision in an IEEE 802.11-based network--namely, SYNC and HN \cite{elsevier14}. Notice that the geometry for a SYNC is limited to the tagged vehicle's carrier-sense range because the packet collision occurs among the vehicles who are sensed but cannot avoid a collision due to allocation of a same backoff value \cite{arxiv19}. On the other hand, the geometry for a HN is greater than a SYNC because the vehicles causing this type of packet collision are placed outside of the tagged vehicle's carrier-sense range \cite{arxiv19}, which figuratively defines ``hidden nodes.''

\begin{lemma}\label{lemma_Psync}
(Probability of a SYNC). \textit{The probability that a packet collides over the air due to a synchronized collision is given by}
\begin{align}\label{eq_Psync}
&\mathsf{P}_{\text{sync}} \left(n_{cs}, \text{CW} \right)\nonumber\\
&= 1 - \sum_{n_{tx} = 0}^{n_{cs}} \Bigg[ \big[n_{tx} \le \text{CW}\big] \left(\begin{array}{c}n_{cs} \vspace{0.06 in}\\ n_{tx}\end{array}\right) \tau^{n_{tx}} \left(1 -\tau\right)^{n_{cs} - n_{tx}}\nonumber\\
&\hspace{1.8 in}\frac{\text{CW}!}{\left(\text{CW}-n_{tx}\right)! \hspace{0.04 in} \text{CW}^{n_{tx}}} \Bigg]
\end{align}
\end{lemma}

\textit{Proof:} Let us define the probability of a SYNC as
\begin{align}\label{eq_Psync_proof}
\mathsf{P}_{\text{sync}} \left(n_{cs}, \text{CW} \right) &= 1 - \mathbb{P}\left[ \text{No SYNC} \hspace{0.05 in} | \hspace{0.05 in} n_{cs} > 0\right].
\end{align}
Assuming independence between two events forming the condition relationship in (\ref{eq_Psync_proof}), we can rewrite as
\begin{align}\label{eq_Pnosync}
&\mathbb{P}\left[ \text{No SYNC} \hspace{0.05 in} | \hspace{0.05 in} n_{cs} > 0 \right]\nonumber\\
&\hspace{0.23 in}= \mathbb{P}\left[ \text{No SYNC} \right] \mathbb{P}\left[ n_{cs} > 0 \right]\nonumber\\
&\hspace{0.2 in}\stackrel{(a)}{\approx} \mathbb{P}\left[ \text{No SYNC} \right]\nonumber\\
&\hspace{0.23 in}= \sum_{n_{tx} = 0}^{n_{cs}} \Big( \big[n_{tx} \le \text{CW}\big] \mathbb{P}\left[ n_{tx} \text{ Tx's} \right]\nonumber\\
&\hspace{0.8 in} \mathbb{P}\left[ \text{All different backoffs among them} \right] \Big)
\end{align}
where $\left[n_{tx} < \text{CW}\right]$ gives an Iverson bracket:
\begin{align}
\big[n_{tx} \le \text{CW}\big] &= \begin{cases}1, \hspace{0.1 in} n_{tx} \le \text{CW}\vspace{0.03 in}\\0, \hspace{0.1 in}\text{otherwise}\end{cases}
\end{align}
It indicates that in order for a SYNC not to occur, there must be only $n_{tx} \le$ CW STAs transmitting, which leaves a possibility that all STAs are assigned all different backoff values. In other words, a SYNC can never be avoided when $n_{tx} > \text{CW}$ STAs compete for the medium.

Step (a) comes from the following rationale. A key condition for a point process to be a PPP is that the number of points falling in a bounded Borel set $\mathcal{A}$ is a Poisson random variable with the parameter of $\lambda |\mathcal{A}|$ \cite{haenggi05} where $|\mathcal{A}|$ denotes the area of an arbitrary two-dimensional space $\mathcal{A}$. Refer to Figs. 6 and 7 of \cite{arxiv19} that $\mathsf{A}_{\text{col}}$ forms a circular space in which a point $x$ is located at the origin of the center and another point $y$ is placed $r$ away from the origin. That is, $|\mathsf{A}_{\text{col}}| = \pi r_{cs}^2$. Based on this, we can exploit the CDF of the distance $\mathsf{l}$ between the two arbitrary points for calculation of $1 - \mathbb{P} \left( \text{No other node in } \mathsf{A}_{\text{col}} \right)$, which is written as
\begin{align}\label{eq_first}
\mathbb{P} \left( n_{cs} > 0 \right) &\stackrel{(b)}{=} \mathbb{P}(n > 0, {\rm{~}} \mathsf{A}_{\text{col}})\nonumber\\
&{\rm{~}}= 1 - \mathbb{P}(n_{cs} = 0, \mathsf{A}_{\text{col}})\nonumber\\
&{\rm{~}}= 1 - \frac{\left( \lambda \pi r_{cs}^2 \right)^0 e^{-\lambda \pi r_{cs}^2}}{0!}\nonumber\\
&{\rm{~}}= 1 - e^{-\lambda \pi r_{cs}^2}, {\rm{~~}} r_{cs} \ge 0.
\end{align}
For (b), we assume the existence of two nodes at least: one for $\mathsf{vT}$ and the other as a potential SYNC-causing node. Here, it is important to note that with a realistic value for $r_{cs}$, $\mathbb{P} \left( n_{cs} > 0 \right) \approx 1$. This justifies approximation (a) in (\ref{eq_Pnosync}).

Now, we present specifics for each term of (\ref{eq_Pnosync}). It is worth to note from the summation in (\ref{eq_Pnosync}) that we consider every possible number of actually transmitting STAs (i.e., $n_{tx}$) among all the STAs sensed by the tagged vehicle's (i.e., $n_{cs}$).

The first term inside the summation of (\ref{eq_Pnosync}) is elaborated as
\begin{align}\label{eq_Psync_proof_term1}
\mathbb{P}\left[ n_{tx} \text{ Tx's} \right] = \left(\begin{array}{c}n_{cs} \vspace{0.06 in}\\ n_{tx}\end{array}\right) \tau^{n_{tx}} \left(1 -\tau\right)^{n_{cs} - n_{tx}}
\end{align}
with $\tau$ as has been given in Proposition \ref{proposition_tau}.

The second term inside the summation in (\ref{eq_Pnosync}) indicates the probability that all the $n_{tx}$ vehicles ($ \forall 0, 1, \cdots, n_{cs}$) are assigned different backoff counter values, which is given by
\begin{align}
&\mathbb{P}\left[ \text{All different backoffs among them} \right]\nonumber\\
&= \frac{\text{All different backoff selections}}{\text{All possible backoff selections by } n_{Tx} \text{ STAs}}\nonumber\\
&= \frac{\text{CW}!\left[\left(\text{CW}-n_{tx}\right)!\right]^{-1}}{\text{CW}^{n_{tx}}}
\end{align}

This yields
\begin{align}\label{eq_Pnosync_proven}
&\mathbb{P}\left[ \text{No SYNC} \hspace{0.05 in} | \hspace{0.05 in} n_{cs} > 0 \right]\nonumber\\
&= \sum_{n_{tx} = 0}^{n_{cs}} \Bigg[ \big[n_{tx} \le \text{CW}\big] \left(\begin{array}{c}n_{cs} \vspace{0.06 in}\\ n_{tx}\end{array}\right) \tau^{n_{tx}} \left(1 -\tau\right)^{n_{cs} - n_{tx}}\nonumber\\
&\hspace{1.8 in}\frac{\text{CW}!}{\left(\text{CW}-n_{tx}\right)! \hspace{0.04 in} \text{CW}^{n_{tx}}} \Bigg]
\end{align}

Therefore, the probability of a SYNC can be formulated as
\begin{align}
&\mathsf{P}_{\text{sync}}\left(n_{cs}, \text{CW} \right)\nonumber\\
&= 1 - \sum_{n_{tx} = 0}^{n_{cs}} \Bigg[ \big[n_{tx} \le \text{CW}\big] \left(\begin{array}{c}n_{cs} \vspace{0.06 in}\\ n_{tx}\end{array}\right) \tau^{n_{tx}} \left(1 -\tau\right)^{n_{cs} - n_{tx}}\nonumber\\
&\hspace{1.8 in}\frac{\text{CW}!}{\left(\text{CW}-n_{tx}\right)! \hspace{0.04 in} \text{CW}^{n_{tx}}} \Bigg]
\end{align}
which completes the proof.
\hfill$\blacksquare$

\begin{lemma}\label{lemma_Phn}
(Probability of a HN). \textit{The probability that a packet collides over the air due to a synchronized collision is given by}
\begin{align}\label{eq_Phn}
&\mathsf{P}_{\text{hn}} \left(n_{hn}, \text{CW} \right)\nonumber\\
&= 1 - \sum_{n_{tx} = 0}^{n_{cs}} \Bigg[ \big[n_{tx} \le \text{CW}\big] \left(\begin{array}{c}n_{cs} \vspace{0.06 in}\\ n_{tx}\end{array}\right) \tau^{n_{tx}} \left(1 -\tau\right)^{n_{cs} - n_{tx}}\nonumber\\
&\hspace{2 in} \mathbb{E}_{c}\left[ \left(\frac{\left|\mathcal{S}_{\sim\text{hn}}\right|}{\text{CW}}\right)^{n_{tx}} \right] \Bigg]
\end{align}
\end{lemma}

\textit{Proof:} Before starting derivation of an equation, it is significant to discuss the critical difference between a SYNC and a HN. While a SYNC occurs in a single slot, a HN can occur at any time instant within an entire BSM (i.e., $l_{bcn}$ slots), since the colliding STAs are unable to sense each other in a HN situation. Specifically, there are two possible scenarios in which a hidden terminal causes a collision with the tagged vehicle: (i) the tagged vehicle starts sending while a hidden terminal is transmitting, or (ii) a hidden terminal starts sending while the tagged vehicle is transmitting. As such, the key difference is that unlike a SYNC, a HN collision can occur even when the STAs transmitting colliding packets were assigned different backoff values.

We start the formulation in a similar way to $\mathsf{P}_{\text{sync}}$ that was shown in (\ref{eq_Psync_proof}), which is given by
\begin{align}\label{eq_Phn_proof}
\mathsf{P}_{\text{hn}} \left(n_{hn}, \text{CW} \right) &= 1 - \mathbb{P}\left[ \text{No HN} \hspace{0.05 in} | \hspace{0.05 in} n_{cs} > 0\right]
\end{align}
where
\begin{align}\label{eq_PnoHN}
&\mathbb{P}\left[ \text{No HN} \hspace{0.05 in} | \hspace{0.05 in} n_{cs} > 0 \right]\nonumber\\
&\hspace{0.2 in}\stackrel{(a)}{\approx} \mathbb{P}\left[ \text{No HN} \right]\nonumber\\
&\hspace{0.23 in}= \sum_{n_{tx} = 0}^{n_{cs}} \Big( \big[n_{tx} \le \text{CW}\big] \mathbb{P}\left[ n_{tx} \text{ Tx's} \right]\nonumber\\
&\hspace{0.9 in} \mathbb{P}\left[ \text{No hidden terminal transmission} \right] \Big)
\end{align}
Approximation (a) is based on the same rationale as one given in (\ref{eq_Pnosync}).

Also, the same Iverson bracket with one shown (\ref{eq_Pnosync}), i.e., $\big[n_{tx} \le \text{CW}\big]$), is given as the first term inside the summation. That is, if the number of transmitting hidden terminals exceeds the current CW, there is no chance that the tagged vehicle can avoid a HN since the existence of hidden terminals.

The second term has also been derived in (\ref{eq_Psync_proof_term1}). The only difference here is that $n_{cs}$ is replaced with $n_{hn} = 3n_{cs}$ assuming the same density. The reason is that the area of a HN has been found as a donut-shaped region formed between circumferences with the radii of $r_{cs}$ and $2r_{cs}$ as illustrated in Fig. 17 of \cite{arxiv19}.

\begin{figure}
\centering
\includegraphics[width = \linewidth]{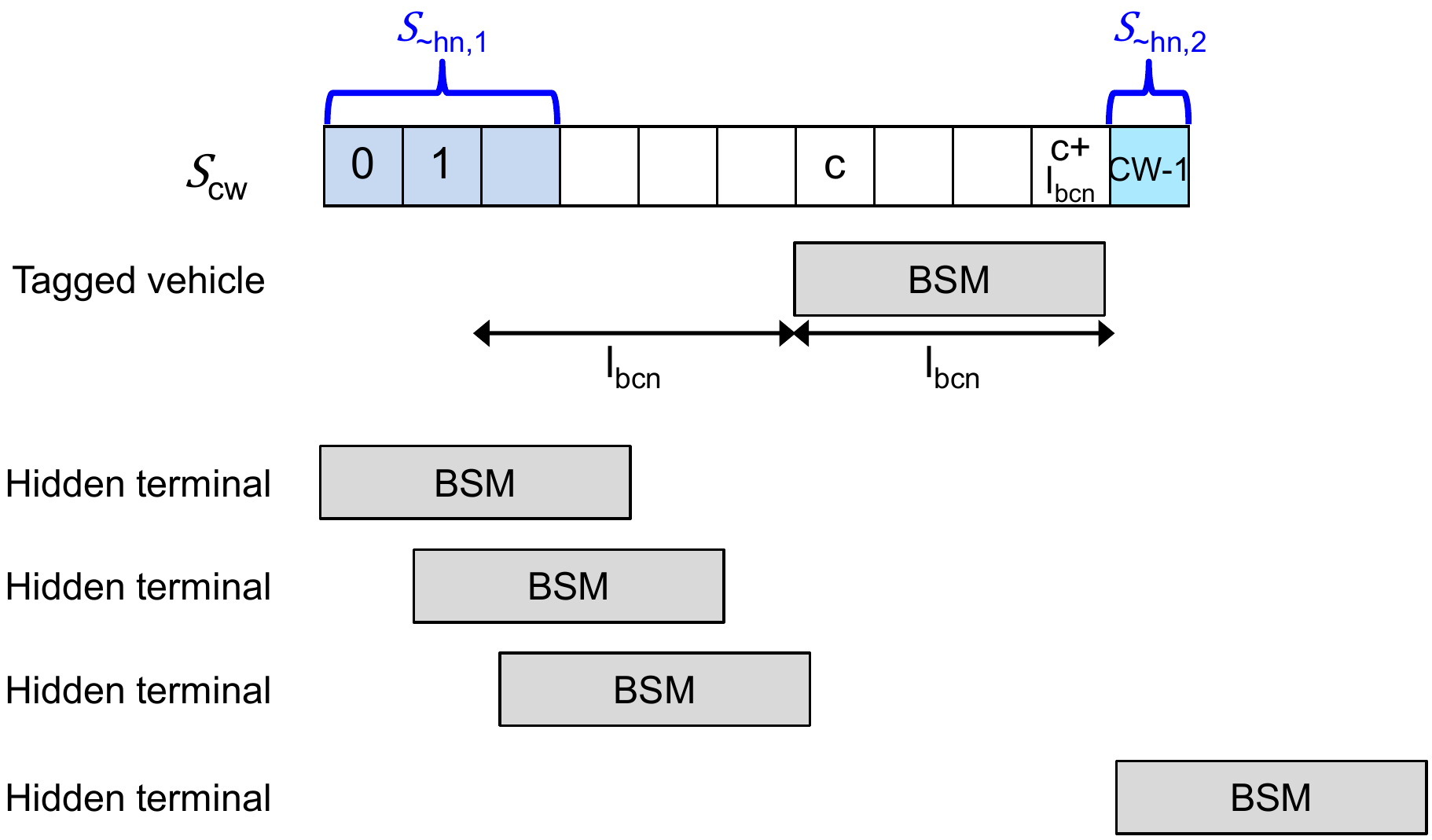}
\caption{An example of $\mathcal{S}_{\sim\text{hn}}$}
\label{fig_nohn}
\end{figure}

The last term in (\ref{eq_PnoHN}) is found as follows. Let $\mathcal{S}_{\text{cw}}$ denote the set of bakcoff values for a given CW, i.e., $\mathcal{S}_{\text{cw}} := \left\{ 0, 1, \cdots, \text{CW}-1 \right\}$. Fig. \ref{fig_nohn} illustrates an example scenario of no HN: a hidden terminal's packet does not overlap with the tagged vehicle's. Sets $\mathcal{S}_{\sim\text{hn},1}$ and $\mathcal{S}_{\sim\text{hn},2}$ contain backoff values (before and after the tagged vehicle's packet, respectively), with which a hidden terminal does not collide with the tagged vehicle, where $\mathcal{S}_{\sim\text{hn},1}, \mathcal{S}_{\sim\text{hn},2} \subset \mathcal{S}_{\text{cw}}$. That way, we define the set of all possible backoff values for a hidden terminal without causing a HN as $\mathcal{S}_{\sim\text{hn}} = \mathcal{S}_{\sim\text{hn},1} \cup \mathcal{S}_{\sim\text{hn},2}$. Based on the definition, we now can derive the term that we are trying to find as
\begin{align}
\mathbb{P}\left[ \text{No hidden terminal transmission} \right] = \mathbb{E}_{c}\left[ \left(\frac{\left|\mathcal{S}_{\sim\text{hn}}\right|}{\text{CW}}\right)^{n_{tx}} \right]
\end{align}
It is critical to note that $\mathcal{S}_{\sim\text{hn}}$ depends on which backoff value $c$ the tagged vehicle is assigned, as shown in Fig. \ref{fig_nohn}. This is the reason that an average $\mathbb{E}\left[\cdot\right]_{c}$ is needed.

This leads to
\begin{align}
&\mathbb{P}\left[ \text{No HN} \hspace{0.05 in} | \hspace{0.05 in} n_{cs} > 0 \right]\nonumber\\
&= \sum_{n_{tx} = 0}^{n_{cs}} \Bigg[ \big[n_{tx} \le \text{CW}\big] \left(\begin{array}{c}n_{cs} \vspace{0.06 in}\\ n_{tx}\end{array}\right) \tau^{n_{tx}} \left(1 -\tau\right)^{n_{cs} - n_{tx}}\nonumber\\
&\hspace{2 in} \mathbb{E}_{c}\left[ \left(\frac{\left|\mathcal{S}_{\sim\text{hn}}\right|}{\text{CW}}\right)^{n_{tx}} \right] \Bigg]
\end{align}

Therefore, the probability of a HN can be formulated as
\begin{align}
&\mathsf{P}_{\text{hn}}\left(n_{hn}, \text{CW} \right)\nonumber\\
&= 1 - \sum_{n_{tx} = 0}^{n_{cs}} \Bigg[ \big[n_{tx} \le \text{CW}\big] \left(\begin{array}{c}n_{cs} \vspace{0.06 in}\\ n_{tx}\end{array}\right) \tau^{n_{tx}} \left(1 -\tau\right)^{n_{cs} - n_{tx}}\nonumber\\
&\hspace{2 in} \mathbb{E}_{c}\left[ \left(\frac{\left|\mathcal{S}_{\sim\text{hn}}\right|}{\text{CW}}\right)^{n_{tx}} \right] \Bigg]
\end{align}
which completes the proof.
\hfill$\blacksquare$

\begin{theorem}\label{theorem_pdr}
Based on the parameters defined through this paper, the $\mathsf{PDR}$ is formulated as
\begin{align}\label{eq_PDR}
\mathsf{PDR} = \tau \left( 1 - \mathsf{P}_{\text{col}} \right)
\end{align}
where
\begin{align}
\mathsf{P}_{\text{col}} = \mathsf{P}_{\text{sync}} \left(1 - \mathsf{P}_{\text{hn}}\right) + \mathsf{P}_{\text{hn}}\left(1 - \mathsf{P}_{\text{sync}}\right) + \mathsf{P}_{\text{sync}}\mathsf{P}_{\text{hn}}
\end{align}
\end{theorem}

\textit{Proof:}
\begin{align}\label{eq_PDR_proof}
\mathsf{PDR} &= \mathbb{P}\left[\text{No collision over the air } | \text{ Packet transmission} \right]\nonumber\\
&\stackrel{(a)}{=} \mathsf{P}_{\text{col}} \tau
\end{align}
where (a) follows from the assumption that the two events, ``No collision over the air'' and ``Packet transmission,'' are independent.

Also, notice that $\mathsf{P}_{\text{col}}$ gives the probability of collision which is formulated as
\begin{align}\label{eq_Pcol}
&\mathsf{P}_{\text{col}}\nonumber\\
&= \mathbb{P}\left[\text{SYNC only}\right] + \mathbb{P}\left[\text{HN only}\right] + \mathbb{P}\left[\text{Both SYNC and HN}\right]\nonumber\\
&= \mathsf{P}_{\text{sync}} \left(1 - \mathsf{P}_{\text{hn}}\right) + \mathsf{P}_{\text{hn}}\left(1 - \mathsf{P}_{\text{sync}}\right) + \mathsf{P}_{\text{sync}}\mathsf{P}_{\text{hn}}
\end{align}
Substituting (\ref{eq_Pcol}) into (\ref{eq_PDR_proof}) completes the proof.
\hfill$\blacksquare$

\begin{figure}
\centering
\includegraphics[width = \linewidth]{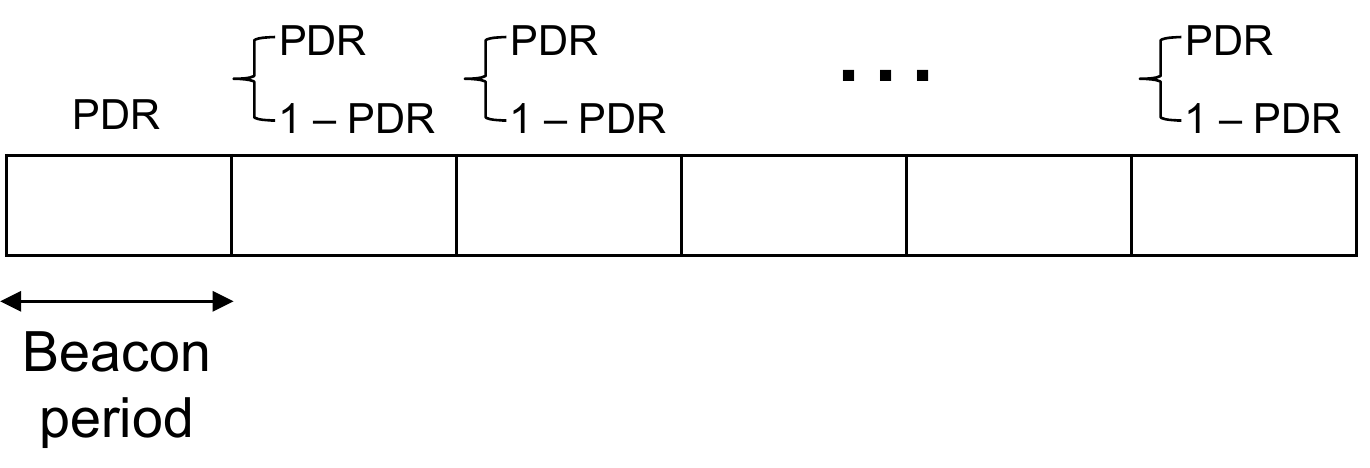}
\caption{Formulation of $\mathsf{IRT}$ as a geometric distribution}
\label{fig_irt_illust}
\end{figure}

\subsection{$\mathsf{IRT}$}
The second metric through which the performance of the proposed protocol is evaluated is the $\mathsf{IRT}$. Fig. \ref{fig_irt_illust} illustrates the formulation of an $\mathsf{IRT}$ as a geometric distribution.

Notice that the unit of a quantity of $\mathsf{IRT}$ is ``the number of slots.'' As such, one needs to multiply a slot time when needing to display an $\mathsf{IRT}$ in the unit of time (i.e., seconds).

\begin{lemma}\label{lemma_IRT}
(Distribution of $\mathsf{IRT}$). \textit{As aforementioned, the probability that an $\mathsf{IRT}$ occurs in the next $k$th beaconing period after the last successful delivery follows a geometric distribution, which can be formally written as}
\begin{align}
\mathsf{IRT} = \left(1 - \mathsf{PDR}\right)^{\nu-1} \mathsf{PDR}
\end{align}
\textit{where $\nu$ denotes the number of unsuccessful receptions.}
\end{lemma}

\textit{Proof:} For occurrence of an ``$\mathsf{IRT}$,'' we suppose to start from a successful reception, then measure how many beaconing periods are needed until the next successful reception. This is the reason that in Fig. \ref{fig_irt_illust}, the first beaconing period is set to have the probability of $\mathsf{PDR}$, and thereafter the possibility is left open between $\mathsf{PDR}$ and $1 - \mathsf{PDR}$. It is formally written as
\begin{align}
\mathsf{IRT} &\sim \text{Geo}\left(\mathsf{PDR}\right)\nonumber\\
&= \mathbb{P}\left[\mathsf{IRT} = \nu\right]\nonumber\\
&= \left(1 - \mathsf{PDR}\right)^{\nu-1} \mathsf{PDR}
\end{align}
\hfill$\blacksquare$

\section{Results and Discussions}\label{sec_results}
This section verifies the accuracy of the analysis presented in Section \ref{sec_stochastic}, by comparing to Monte Carlo simulations of the network defined in Section \ref{sec_system_model}.

\begin{figure}
\centering
\includegraphics[width = \linewidth]{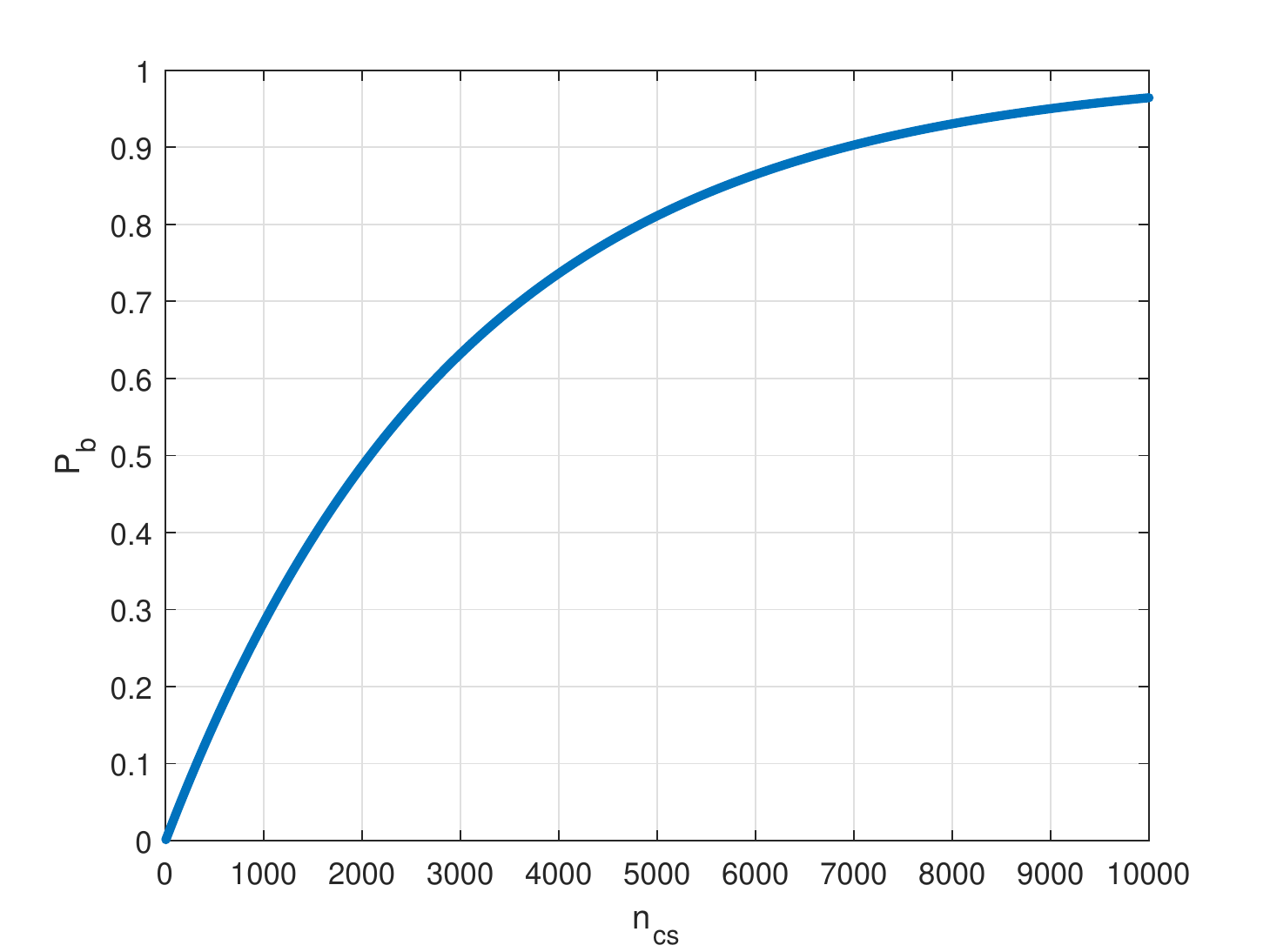}
\caption{$\mathsf{P}_{b}$ versus $n_{cs}$}
\label{fig_Pb_uniform_vs_ncs}
\end{figure}

\begin{figure}
\centering
\includegraphics[width = \linewidth]{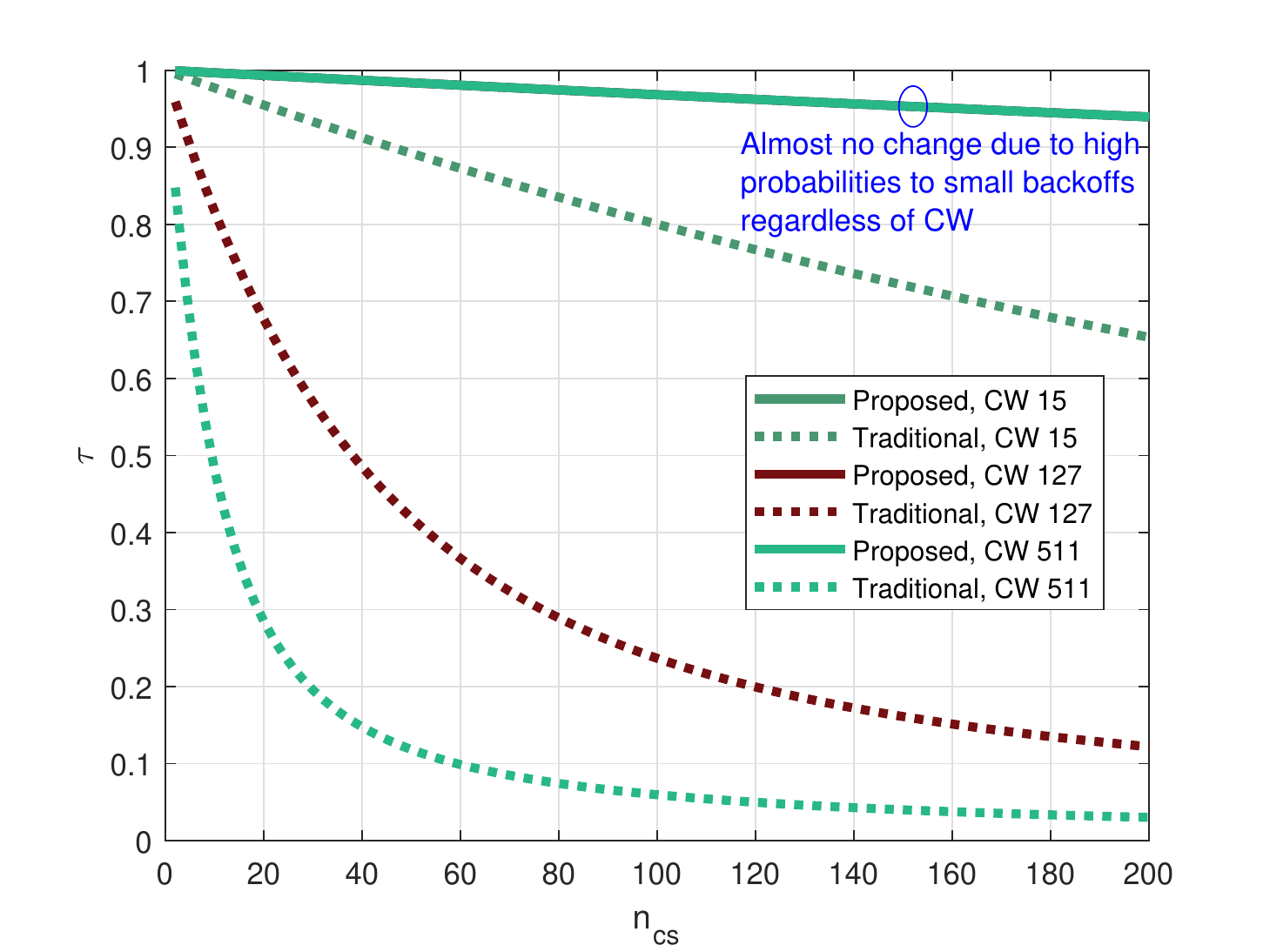}
\caption{$\tau$ versus $n_{cs}$}
\label{fig_tau}
\end{figure}

\subsection{Results}

\subsubsection{$\mathsf{P}_{b}$}
Fig. \ref{fig_Pb_uniform_vs_ncs} demonstrates the $\mathsf{P}_{b}$ that has been resulted from the ``uniform'' approximated $\tau$. Overall, the probability that a certain slot is found busy with another node can be kept relatively low--i.e., around 15\% even with 500 competing nodes within one's carrier-sense range. The reason is the small value for $\tau$ as shown in Fig. \ref{fig_tau}, which is plausible because a beaconing period contains a very large number of slots--i.e., $L_{bcn}$ is as large as 750 with 50 msec and 66.7 $\mu$sec for a beaconing period and a slot time, respectively. The numbers $L_{bcn} =750$ and time length of 50 msec come from the fact that a CCH takes 50 msec of an entire 100-msec beaconing period, based on IEEE 1609.4 that defines the ``alternation'' between CCH and SCH \cite{ieee16094}.

Also, the results match one's intuitions: $\mathsf{P}_{b}$ is increased with a greater $n_{cs}$. One can observe that a smaller value of $r$ yields a lower $\mathsf{P}_{b}$ (which, in turn, leads to a higher $\mathsf{P}_{\text{start}}$ as such). The rationale behind this result is that a smaller $r$ serves as a more drastic decrement of $\mathsf{P}_{\text{ini}}$ as shown in (\ref{eq_Pini}).

\subsubsection{$\tau$}
Fig. \ref{fig_tau} shows the probability that a vehicle has been able to go through a backoff process and transmit a BSM, denoted by $\tau$, versus the number of vehicles within its carrier-sense range. Notice that the key change from the reference \cite{arxiv19} is the Markov chain due to proposition of $\mathsf{P}_{\text{ini}}$.

\begin{figure}
\centering
\includegraphics[width = \linewidth]{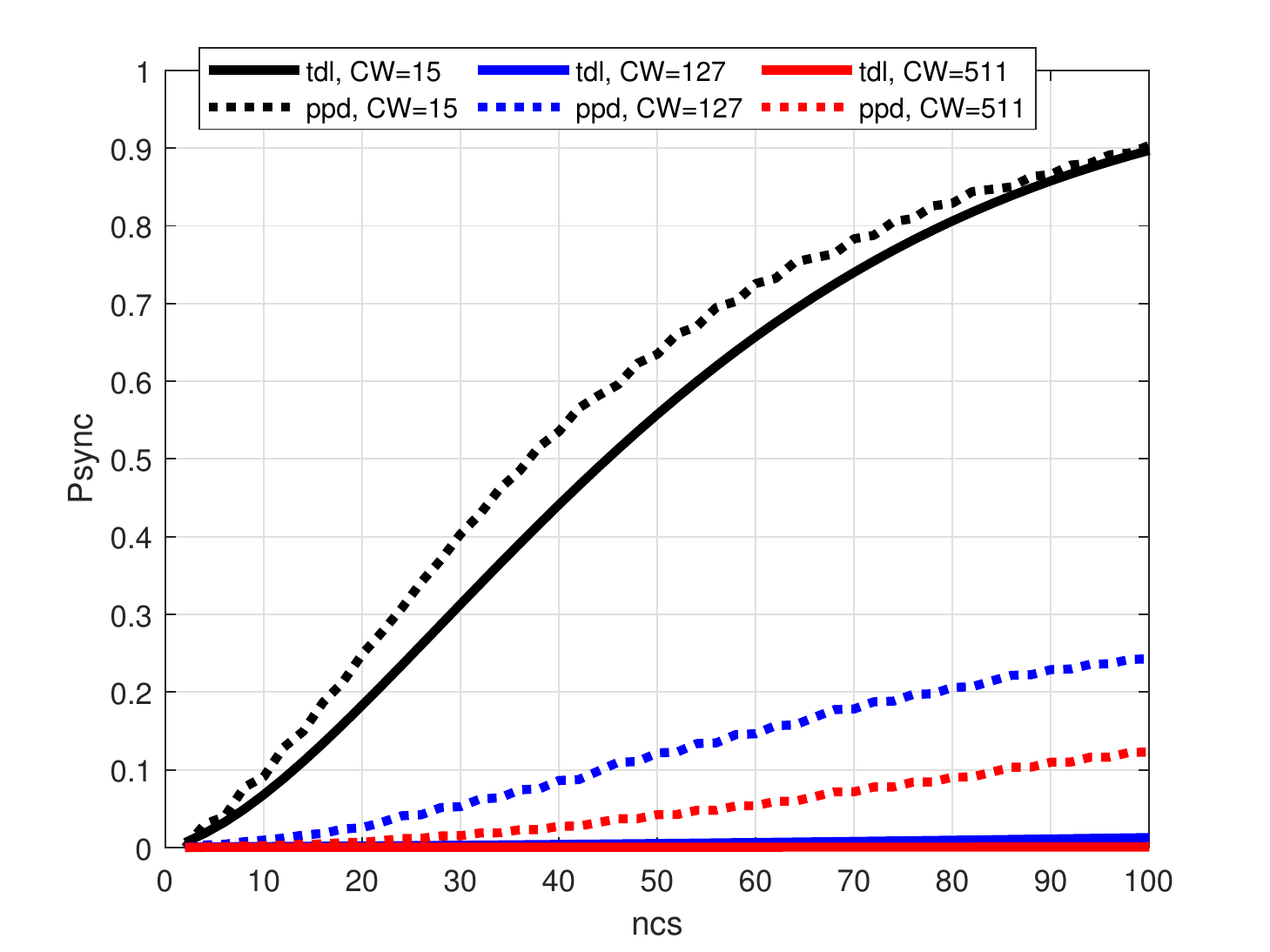}
\caption{$\mathsf{P}_{\text{sync}}$ versus $n_{cs}$}
\label{fig_Psync}
\end{figure}
\begin{figure}
\centering
\includegraphics[width = \linewidth]{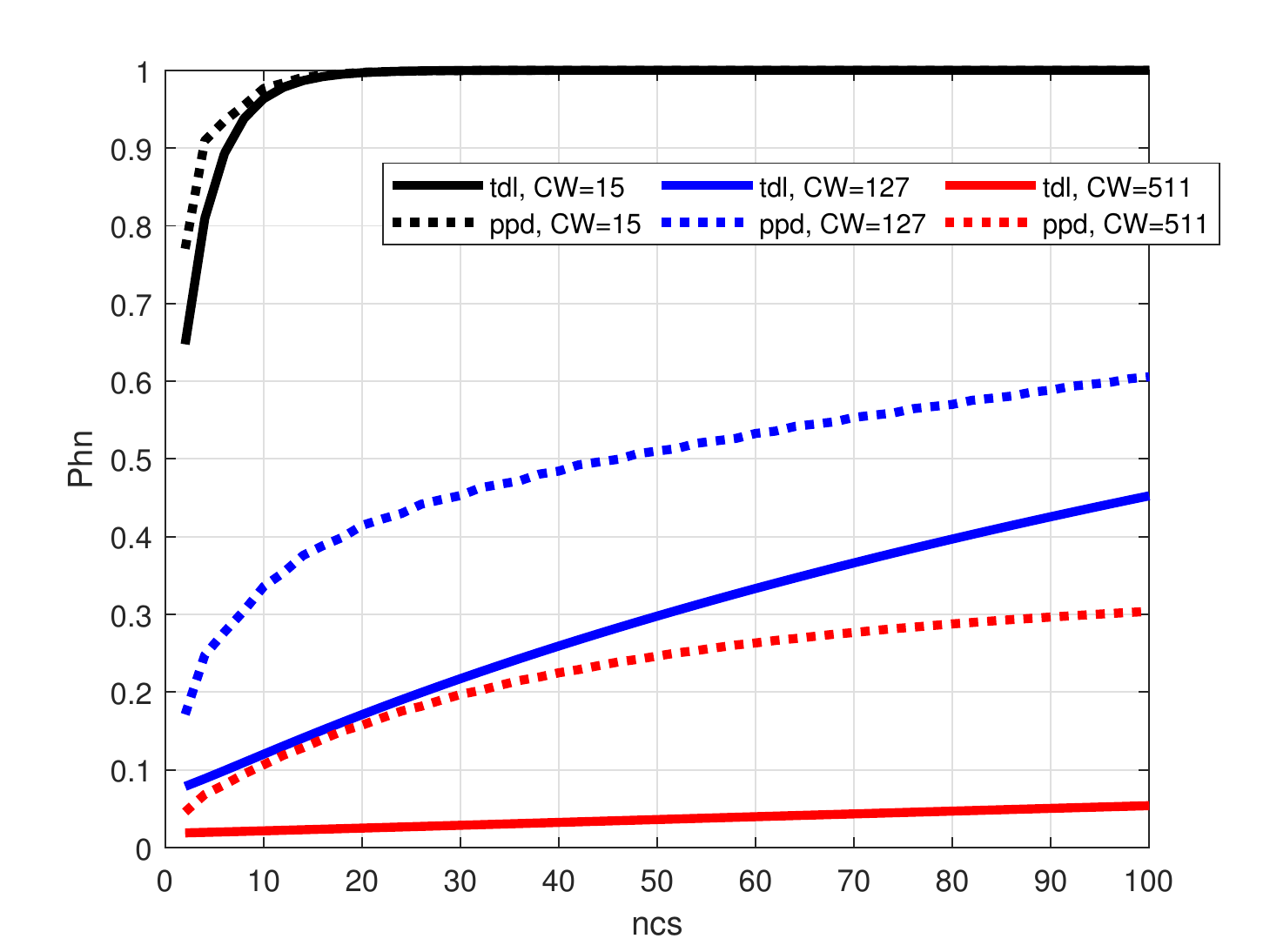}
\caption{$\mathsf{P}_{\text{hn}}$ versus $n_{cs}$}
\label{fig_Phn}
\end{figure}
\begin{figure}
\centering
\includegraphics[width = \linewidth]{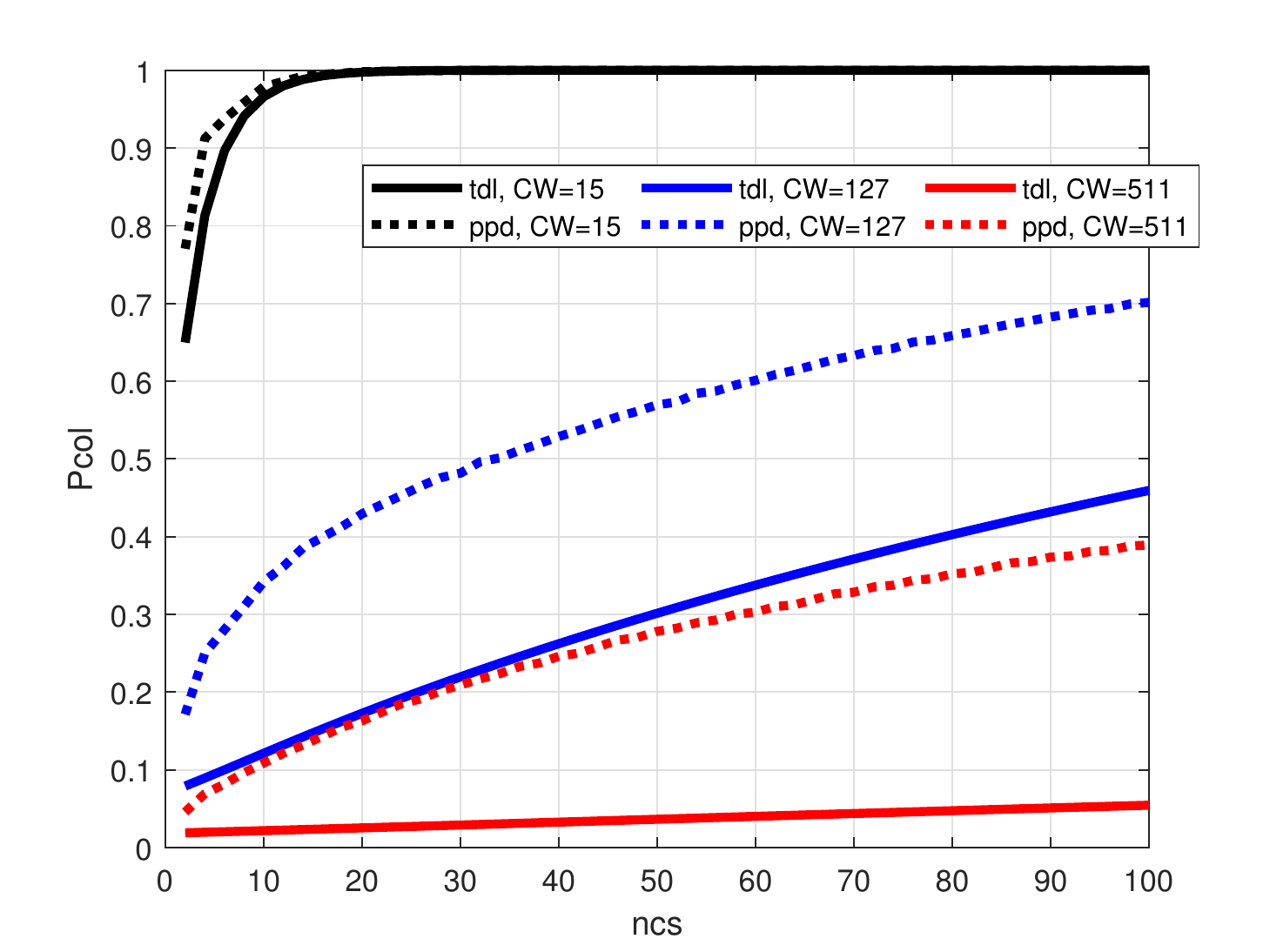}
\caption{$\mathsf{P}_{\text{col}}$ versus $n_{cs}$}
\label{fig_Pcol}
\end{figure}

\subsubsection{$\mathsf{P}_{\text{col}}$}\label{sec_results_pcol}

Figs. \ref{fig_Psync} and \ref{fig_Phn} demonstrate the probabilities of collision by a SYNC and HN, respectively, while Fig. \ref{fig_Pcol} presents the resulting probability of a collision, which was given in (\ref{eq_Pcol}). The probabilities are displayed versus $n_{cs}$ according to the backoff function type and CW. There are several important points to discuss from Figs. \ref{fig_Psync} and \ref{fig_Phn}: (i) a larger CW is effective in alleviating both SYNC and HN, and therefore $\mathsf{P}_{\text{col}}$ as a direct consequence; (ii) For a CW, $\mathsf{P}_{\text{hn}} > \mathsf{P}_{\text{sync}}$ with a node density. The reasoning is that the number of hidden terminals is larger than that of STAs causing a SYNC since the area of HN is larger than that of SYNC \cite{arxiv19}; (iii) the proposed protocol yields a higher $\mathsf{P}_{\text{col}}$ than the traditional CSMA/CA since, on average, it allows certain nodes more likely to transmit.

\begin{figure}
\centering
\includegraphics[width = \linewidth]{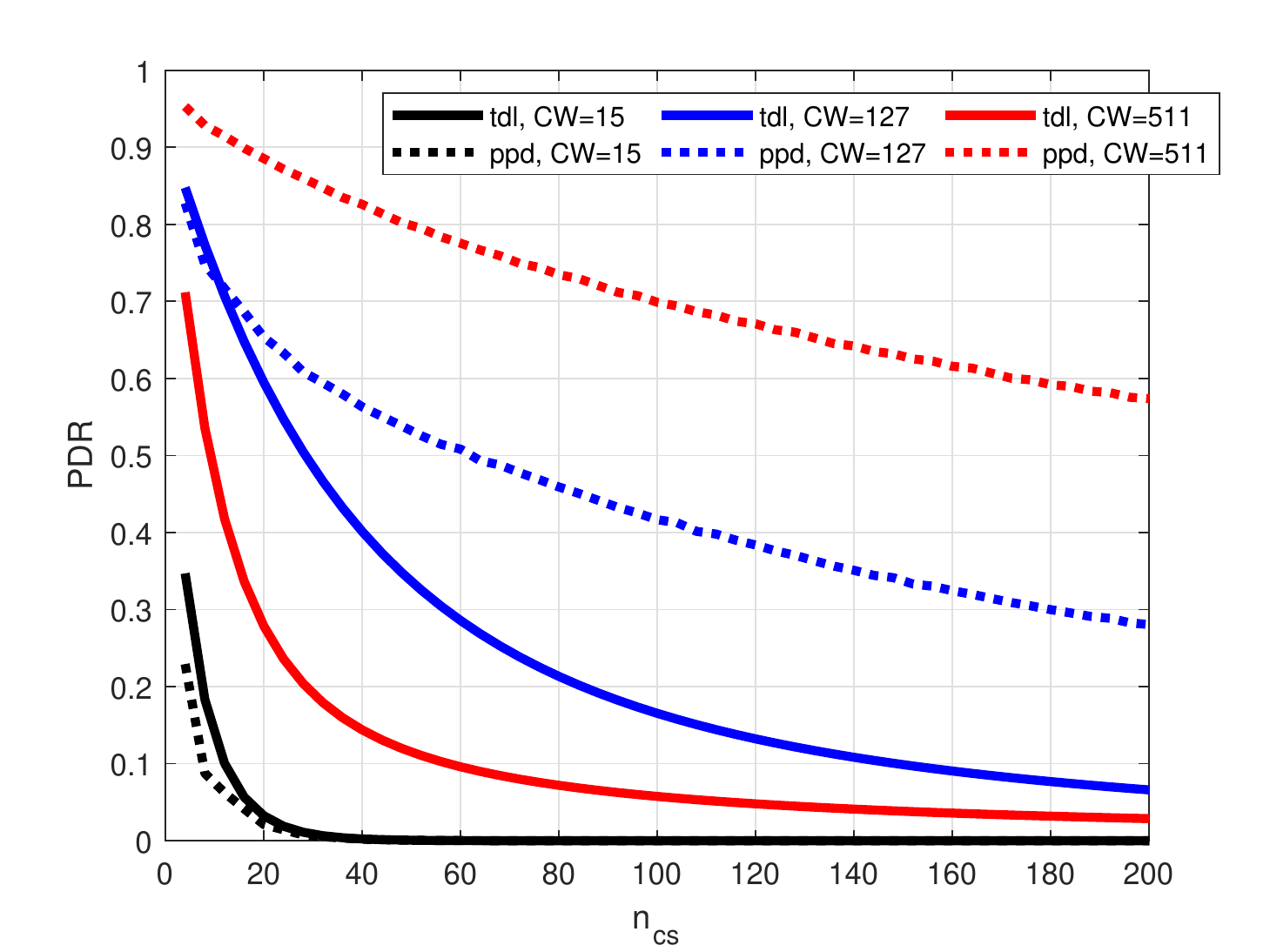}
\caption{$\mathsf{PDR}$ versus $n_{cs}$}
\label{fig_PDR}
\end{figure}

\subsubsection{$\mathsf{PDR}$}
Fig. \ref{fig_PDR} shows $\mathsf{PDR}$ versus $n_{cs}$. We note two key observations as follows. First, the proposed protocol yields a higher PDR while it showed an unfavorable $\mathsf{P}_{\text{col}}$, the reason of which is interpreted as a higher margin in $\tau$. Second, the margin that the proposed protocol yields in $\mathsf{PDR}$ gets increased with a larger CW. The reason is that $\mathsf{P}_{\text{col}}$ gets very small (i.e., $\approx 0$) with a larger CW such as 511.

\begin{figure}
\centering
\begin{subfigure}{.45\textwidth}
\centering
\includegraphics[width = \linewidth]{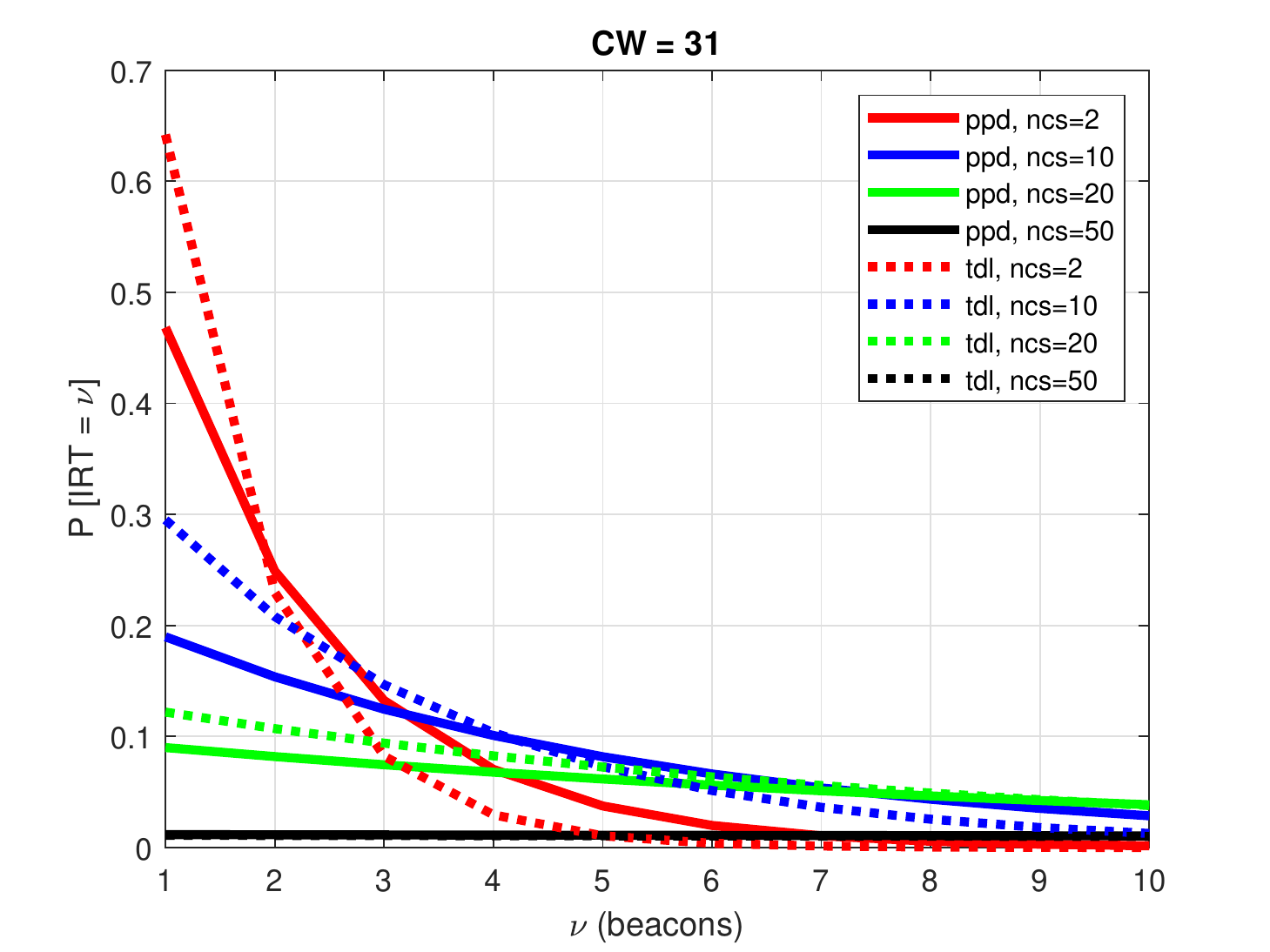}
\caption{CW = 31}
\label{fig_IRT_CW31}
\end{subfigure}
\begin{subfigure}{.45\textwidth}
\centering
\includegraphics[width = \linewidth]{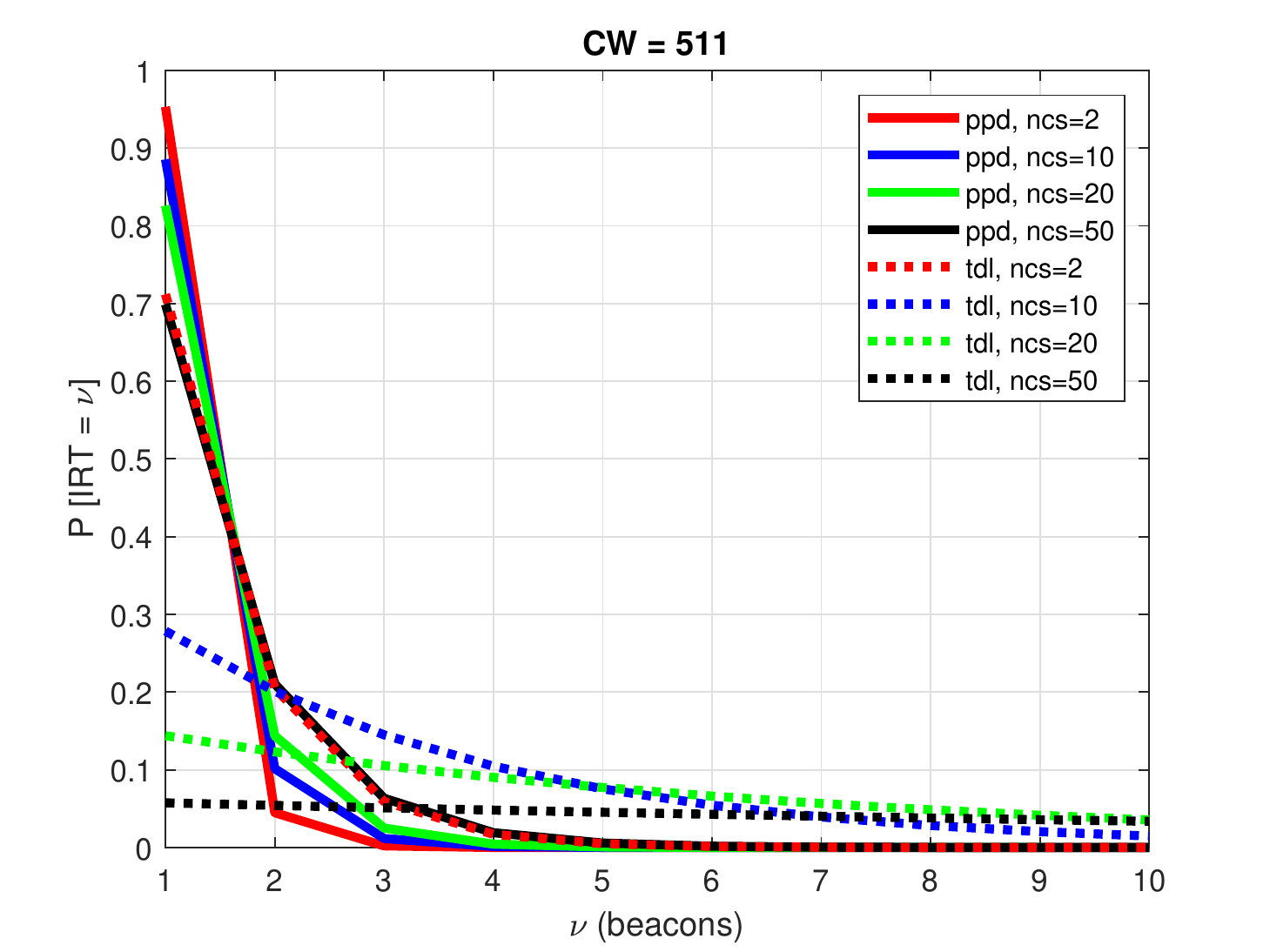}
\caption{CW = 511}
\label{fig_IRT_CW511}
\end{subfigure}
\caption{$\mathsf{IRT}$ versus $\nu$}
\label{fig_IRT}
\end{figure}

\subsubsection{$\mathsf{IRT}$}
Fig. \ref{fig_IRT} shows the probability mass function (PMF) of $\mathsf{IRT}$ with respect to $\nu$, the number of slots until the next successful packet delivery. Comparing \ref{fig_IRT_CW31} and \ref{fig_IRT_CW511} suggests that a larger CW results in a higher probability of smaller $\mathsf{IRT}$s, which is attributed to a higher $\mathsf{PDR}$ with a larger CW as shown in Fig. \ref{fig_PDR}. Within each of \ref{fig_IRT_CW31} and \ref{fig_IRT_CW511}, it is observed that a smaller $n_{cs}$ yields a higher probability of smaller $\mathsf{IRT}$s. The proposed protocol underperforms the traditional CSMA with CW = 31 as seen in Fig. \ref{fig_IRT_CW31} because of the same tendency in $\mathsf{PDR}$. In contrast, with a large CW (i.e., CW = 511), the outperformance of the proposed protocol gets significant with all given values of $n_{cs}$.

\subsection{Discussions}
The proposed scheme yields a higher performance compared to the proposed mechanism in terms of $\mathsf{PDR}$ and $\mathsf{IRT}$ as shown in Figs. \ref{fig_PDR} and \ref{fig_IRT}, respectively. The reason of a higher $\mathsf{PDR}$ achieved by the proposed scheme, despite a higher $\mathsf{P}_{\text{col}}$ as found in Fig. \ref{fig_Pcol}, is attributed to a higher $\tau$ as Fig. \ref{fig_tau} presents.

The predominance of CW in determination of $\mathsf{PDR}$ and $\mathsf{IRT}$ suggests that $\mathsf{P}_{\text{col}}$ acts as the most critical factor in inducing the two quantities. As observed in Fig. \ref{fig_Pcol}, a larger CW yields a lower $\mathsf{P}_{\text{col}}$ and broadens the outperformance of the proposed scheme over the proposed one. Henceforth, all these results suggest that a DSRC network can actively increases CW as an effort to combat a lower $\mathsf{PDR}$ as a result of high road traffic, i.e., a large $n_{cs}$.

\section{Conclusions}
This paper proposed a backoff allocation scheme that is adaptive to the level of accident risk to which a vehicle is exposed. The proposed scheme requires a minimal modification of the current CSMA mechanism adopted for IEEE 802.11p, which can achieve higher generality and applicability to practice. We first identified the driving speed as a key factor determining a crash risk. For analysis, we defined the variance of speed as the key metric and presented closed-form expressions for the probability distribution of the metric. Then, we modeled the IEEE 802.11p CSMA as a Markov process, from which further performance evaluation metrics were derived: i.e., probabilities of transmission within a beaconing period, transmission within a certain slot, collision by a SYNC, collision by a HN, and $\mathsf{IRT}$. From the analysis and numerical results, the main finding was that the performance of a BSM broadcast in an IEEE 802.11p network is predominated by contention rather than collision.

As such, this work has many possible extensions. The key merit of this work is identification of a factor that is critical in causing an accident among the ones that can be obtained on a vehicle autonomously without any external support from infrastructure (e.g., RSU), which fits the ``distributed'' nature of a V2X network. One possible extension is to examine the feasibility of the proposed mechanism for a completely distributed consensus algorithm (e.g., practical Byzantine fault tolerance) for blockchain. Considering that the decentralization is being regarded as one of the most critical factors in the blockchain, the complete decentralization achieved by the proposed scheme will form a critical basis for establishing analytical frameworks for such blockchain-related extension.


\end{document}